\documentclass[useAMS,usenatbib]{mnras}
\usepackage{amsmath,graphicx,bm,amsbsy,amssymb,aas_macros,color,enumerate,lgrind,cmap,rotating,url,array,hyperref,appendix,ctable,capt-of,setspace, float, booktabs}
\usepackage{pdflscape}
\usepackage{natbib}
\newcommand{\refeq}[1]{Eq.~\eqref{#1}}
\newcommand{\refsec}[1]{Section \ref{#1}}

\newcommand{\reffig}[1]{Fig.~\ref{#1}}
\newcommand{\mPsi}{\mathbf{\Psi}}
\newcommand{\mLambda}{\mathbf{\Lambda}}

\newcommand{\C}{\mathbf{C}}

\newcommand{\B}{\mathbf{B}}

\newcommand{\D}{\mathbf{D}}

\newcommand{\beq}{\begin{equation}}
\newcommand{\eeq}{\end{equation}}

\newcommand{\M}{\mathbf{M}}

\def\spose#1{\hbox to 0pt{#1\hss}}
\def\simlt{\mathrel{\spose{\lower 3pt\hbox{$\mathchar"218$}}
   \raise 2.0pt\hbox{$\mathchar"13C$}}}
\def\simgt{\mathrel{\spose{\lower 3pt\hbox{$\mathchar"218$}}
     \raise 2.0pt\hbox{$\mathchar"13E$}}}
 \def\simpropto{\mathrel{\spose{\lower 3pt\hbox{$\mathchar"218$}}
     \raise 2.0pt\hbox{$\propto$}}}

\newcommand{\GHz}{\text{ GHz}}

\newcommand{\be}{\begin{equation}}
\newcommand{\ee}{\end{equation}}

\def\Let@{\def\\{\notag\math@cr}}

\newcolumntype{L}[1]{>{\raggedright\let\newline\\\arraybackslash\hspace{0pt}}m{#1}}
\newcolumntype{C}[1]{>{\centering\let\newline\\\arraybackslash\hspace{0pt}}m{#1}}
\newcolumntype{R}[1]{>{\raggedleft\let\newline\\\arraybackslash\hspace{0pt}}m{#1}}

\newcount\colveccount
\newcommand*\colvec[1]{
        \global\colveccount#1
        \begin{pmatrix}
        \colvecnext
}
\def\colvecnext#1{
        #1
        \global\advance\colveccount-1
        \ifnum\colveccount>0
                \\
                \expandafter\colvecnext
        \else
                \end{pmatrix}
        \fi
}

\title[Improved Model of Diffuse Galactic Radio Emission]{An Improved Model of Diffuse Galactic Radio Emission from 10\,MHz to 5\,THz}
\author[H.~Zheng, et al.]{H.~Zheng$^{1}$\thanks{E-mail:		
		jeff.h.zheng@gmail.com}, M.~Tegmark$^{1}$, J.~S.~Dillon$^{1,2}$, D.A.~Kim$^{2}$, A.~Liu$^{2}$\thanks{Hubble Fellow}, A.~R.~Neben$^{1}$,
	\newauthor J.~Jonas$^{3}$, P.~Reich$^{4}$, W.~Reich$^{4}$\\
	$^{1}$Dept.~of Physics and MIT Kavli Institute, Massachusetts Institute of Technology, 77 Massachusetts Ave., Cambridge, MA 02139, USA\\
	$^{2}$Dept.~of Astronomy and Radio Astronomy Lab, University of California, Berkeley, CA 94720, USA\\
	$^{3}$Dept.~of Physics \& Electronics, Rhodes University, Grahamstown 6140, South Africa\\
	$^{4}$Max-Planck-Institut f\"ur Radioastronomie, Auf dem H\"ugel 69, D-53121 Bonn, Germany\\
}
\date{\today}

\begin{document}
	
\maketitle

\begin{abstract}
	We present an improved Global Sky Model (GSM) of diffuse Galactic radio emission from 10\,MHz to 5\,THz, whose uses include foreground modeling for CMB and 21\,cm cosmology. Our model improves on past work both algorithmically and by adding new data sets such as the \textit{Planck} maps and the enhanced Haslam map. Our method generalises the Principal Component Analysis approach to handle non-overlapping regions, enabling the inclusion of 29 sky maps with no region of the sky common to all. We also perform a blind separation of our GSM into physical components with a method that makes no assumptions about physical emission mechanisms (synchrotron, free-free, dust, etc). Remarkably, this blind method automatically finds five components that have previously only been found ``by hand'', which we identify with synchrotron, free-free, cold dust, warm dust, and the CMB anisotropy. Computing the cross-power spectrum between these blindly extracted components and \emph{Planck} component maps, we find a strong correlation at all angular scales. The improved GSM is available online at \url{http://github.com/jeffzhen/gsm2016}.
\end{abstract} 


%%%%%%%%%%%%%%%%%%%%%%%%%%%%%
%%%%%%%%%%%%%%%%%%%%%%%%%%%%%
%%%%%%%%%%%%%%%%%%%%%%%%%%%%%
\section{Introduction}
Modeling diffuse Galactic radio emission has received great interest in cosmology. Cosmological signals, such as the cosmic microwave background (CMB) and redshifted 21\,cm emission, have to pass through our Galaxy before reaching us, so they are inevitably mixed with foreground contamination from Galactic emission. For CMB experiments, the unpolarised CMB signal dominates over foregrounds in out-of-plane directions, though the situation for polarised CMB signal is about one to two orders of magnitude worse (for reviews, see \citet{all_sorts_of_cmb_papers} and references therein). For the redshifted 21\,cm signals from the Epoch of Reionization (EoR), the foregrounds are thought to be four or more orders of magnitudes higher than the 21\,cm signal (see \citealt{FurlanettoReview,miguelreview} for reviews, and \citealt{MWAJosh, PAPERpspec, PAPERpspec2} for the latest 21\,cm power spectrum upper limits), which makes foreground modeling even more important \citep{foreground1, foreground2, foreground3, foreground4, foreground5, foreground6, foreground7, foreground8,foreground9, foreground10, pober_etal2013b, liu_et_al2014a, liu_et_al2014b, dillon_et_al2015}. Among many efforts to model the foreground are the \textit{Planck Sky Model} (PSM; \citealt{PSM}) and the GSM. 

In the original GSM \citep{GSM}, the authors carried out an exhaustive survey of existing sky maps from 10\,MHz to 100\,GHz, and performed a Principal Component Analysis (PCA) on 11 of the highest quality maps. They found that using just the top three principal components could explain between 90\% and 99\% of the variations in all the maps, depending on frequency and sky direction. By interpolating these three components over frequency, the GSM can therefore model the diffuse Galactic emission anywhere between 10\,MHz and 100\,GHz.

The considerable flexibility and power of the GSM software has resulted in its widespread application. Given that foregrounds are not only ubiquitous in cosmological surveys, but also interesting probes of astrophysical phenomena in their own right, it is unsurprising that the GSM has had broad influence within the astrophysics community. 

Astrophysical research using the GSM has ranged, for example, from limits on radio emission from annihilating dark matter (e.g.~\citealt{Spekkens2013}) to searches for non-Gaussianity in Planck CMB maps (e.g.~\citealt{Novaes2015}) and from EoR power spectrum limits (e.g.~\citealt{PAPERpspec2}) to the discovery of a Fast Radio Burst (e.g.~\citealt{Burke-Spolaor}).

However, there are some notable limitations to the GSM. In terms of frequency coverage, there is a lack of direct observation in the EoR frequency range. The closest maps in frequency are a 45\,MHz map with about $3.6^\circ$ resolution \citep{Alvarez1997MRAO,Maeda1999JMUAR} and a 408\,MHz map with $56'$ resolution \citep{Haslam1981, Haslam1982, Remazeilles2015Haslam}, and for the high resolution GSM version, the entire model is locked to the 408\,MHz map. In terms of sky coverage, the PCA algorithm is performed on a rather small region covered by all 11 maps, mostly in out-of-plane directions, so the resulting principal components may suffer from biased frequency dependencies skewed towards that of the small region covered. In terms of accuracy, each of the 11 data sets are normalised beforehand so that they are treated equally, regardless of their relative accuracy. Thus, the model may suffer unnecessarily from noise and systematics in the maps of lesser quality. Lastly, since the PCA produces orthogonal principal components and orthogonal principal maps, these components are not likely to correspond to actual physical processes such as synchrotron or dust emission. Actual emission mechanisms have similar spatial structures (such as strong emission in the Galactic plane) and thus are not mutually orthogonal.

In this work, we present a new GSM-building method that naturally extends the original PCA algorithm. We use PCA as the initial step to obtain crude estimates of the principal components and their corresponding maps, and we iterate between the components and the maps to find the best fit to all of the data available. This method allows us to include 29 sky maps in the frequency range 10\,MHz to 5\,THz (including the Parkes maps at 85\,MHz and 150\,MHz \citealt{Landecker1970Parkes}), which share no completely common sky coverage. This method can be further extended in the future to incorporate uncertainty information from each map into the fitting process. Furthermore, we use a blind component separation technique to recombine the orthogonal components into those that are physically interpretable as different emission mechanisms. Both our blind spectra and blind component maps agree remarkably well with existing physical models, and additionally are consistent with the \emph{Planck} component maps.

The remainder of this paper is structured as follows. In \refsec{secalgorithm}, we describe our improved GSM-building method. In \refsec{secsurvey}, we describe the 29 sky maps included in this work. In \refsec{secresult}, we present the improved GSM from 10\,MHz to 5\,THz with six components in \refsec{subsecorthogonal}, and estimate the predictive accuracy of the new GSM in \refsec{subsecerrorgsm2}. In \refsec{secrecombine}, we use a blind approach to recombine those components into physically meaningful contributions in \refsec{subsecphysical}, and compare our blind spectra and component maps with existing results in the literature in Sec.~\ref{subsecphysicalspectra} and \ref{subsechighres}, along with a quantitative comparison to \emph{Planck} component maps in \refsec{subsecphysdiscussion}.
Finally, in \refsec{secsummarygsm2}, we conclude by summarizing this work and discussing potential future improvements for the GSM.

%%%%%%%%%%%%%%%%%%%%%%%%%%%%%
%%%%%%%%%%%%%%%%%%%%%%%%%%%%%
%%%%%%%%%%%%%%%%%%%%%%%%%%%%%
\section{Iterative Algorithm for Building a GSM}\label{secalgorithm}

%%%%%%%%%%%%%%%%%%%%%%%%%%%%%
%%%%%%%%%%%%%%%%%%%%%%%%%%%%%
\subsection{Framework}\label{subsecframework}
We describe the GSM in the form of two matrices, an $n_\text{pix}\times n_\text{c}$ map matrix $\M$ and an $n_\text{c}\times n_\text{f}$ normalised spectrum matrix $\mathbf{S}$, where $n_\text{pix}$ is the number of pixels in each sky map, $n_\text{c}$ is the number of components, and $n_\text{f}$ is the number of frequencies for which we have maps. Furthermore, we encapsulate all the sky map data into an $n_\text{pix}\times n_\text{f}$ matrix $\D$, where all maps are normalised to the same level at each frequency.\footnote{Due to incomplete and different sky coverages between maps, a naive normalization will not weigh the maps properly. We discuss our normalization in more detail in \refsec{subseciterativegsm2}.} The GSM then models the sky by
\begin{equation}\label{eqMWD}
\D \approx \M\mathbf{S}.
\end{equation}
Thus to construct a GSM is to find the pair of $\M$ and $\mathbf{S}$ that minimises the cost function
\begin{equation}\label{eqcost}
\|\D-\M\mathbf{S}\|^2.
\end{equation}

In general, because both components in the product $\M\mathbf{S}$ are unknown, we have degeneracies in the form of an invertible $n_\text{c}\times n_\text{c}$ matrix $\mPsi$. For any solution $\M$ and $\mathbf{S}$, an alternative solution $\M' = \M\mPsi^{-1}$ and $\mathbf{S}' = \mPsi\mathbf{S}$ will produce an identical prediction for the data $\D$.

To use the GSM to predict sky maps at previously unmeasured frequencies or sky regions, one interpolates at the desired frequency both the normalised spectra in $\mathbf{S}$ and the overall normalization. For the rest of this work, we choose linear interpolation for the normalised spectra over the log of frequency, and linear interpolation for the log of normalization over the log of frequency. Because normalised spectra are interpolated linearly, the choice of degeneracy matrix $\mPsi$ will have no effect on the predictions of the GSM. 

%For the rest of this section, we first use this framework to explain the PCA algorithm in the original GSM in \refsec{subsecpcagsm1}. We then discuss solving \refeq{eqMWD} in \refsec{subseciterativegsm2}, generalizing \refeq{eqMWD} in \refsec{subsecinterferometric}, and optimizing $\mPsi$ in \refsec{subsecphysical}.

%%%%%%%%%%%%%%%%%%%%%%%%%%%%%
%%%%%%%%%%%%%%%%%%%%%%%%%%%%%
\subsection{PCA Algorithm}\label{subsecpcagsm1}
We first describe the original PCA algorithm in \citet{GSM}, before generalizing it to our iterative algorithm. Given the data matrix $\D$, one first performs an eigen-decomposition of $\D^t\D$:
\begin{equation}\label{DDCLC}
\D^t\D = \C^t\mLambda\C,
\end{equation}
where $\D^t$ denotes the transpose of $\D$, $\C$ is an $n_\text{f}\times n_\text{f}$ orthogonal matrix with eigenvectors as its rows, and $\mLambda$ is an $n_\text{f}\times n_\text{f}$ diagonal matrix with eigenvalues on its diagonal. If the sky can be described by $n_\text{c}$ components where $n_\text{c} < n_\text{f}$, then $\mLambda$ has only $n_\text{c}$ non-zero eigenvalues on its diagonal\footnote{In practice, no eigenvalues are perfect zeros, and we will discuss our choice of $n_\text{c}$ in more details in \refsec{subsecorthogonal}.}, so
\begin{equation}\label{DDCLC2}
\D^t\D = \widetilde{\C}^t\widetilde{\mLambda}\widetilde{\C},
\end{equation}
where $\widetilde{\C}$ and $\widetilde{\mLambda}$ are the $n_\text{c}\times n_\text{f}$ and $n_\text{c}\times n_\text{c}$ parts of their non-tilde counterparts corresponding to the non-zero eigenvalues. One then takes the principal components $\widetilde{\C}$ and solves for the best $\M$ that satisfies
\begin{equation}\label{eqMCD}
\M\widetilde{\C} \approx \D.
\end{equation}

Comparing \refeq{eqMWD} and \refeq{eqMCD}, we see that the PCA algorithm obtains one solution to \refeq{eqMWD}, with $\mathbf{S} = \widetilde{\C}$. However, in practice, the sky maps have various sky coverages with potentially very few overlapping pixels, so rather than using the full $\D$ for \refeq{DDCLC}, the original GSM uses a $\D^*$ that consists of a subset of the rows in $\D$ that corresponds to the common pixels covered at all frequencies. Therefore, the process of obtaining $\widetilde{\C}$ is ignoring the majority of information in $\D$, and the solutions are thus not minimizing the cost function in \refeq{eqcost}. Furthermore, if one wishes to include more data sets, such as we do in this work, there are no overlapping regions that cover all of the frequency range, so it is difficult to apply the original PCA algorithm as-is.

%%%%%%%%%%%%%%%%%%%%%%%%%%%%%
%%%%%%%%%%%%%%%%%%%%%%%%%%%%%
\subsection{Iterative Algorithm}\label{subseciterativegsm2}
To overcome the challenges discussed above, we extend the PCA algorithm by iterating on the results obtained from it. We start by temporarily excluding the data sets with smallest sky coverage one by one, until the remaining data sets have more than $5\%$ common sky coverage. We then use the 5\% common pixels to obtain $n_\text{c}$, $\M^{(0)}$, and $\widetilde{\C}$ following the PCA algorithm described in the previous section, where $\M^{(0)}$ denotes our starting $\M$ at the 0th iteration. Once we obtain $\M^{(0)}$, we fold the temporarily excluded frequencies back into $\D$ and start iterating. At the $i$th iteration we compute
\begin{equation}\label{eqWiter}
\mathbf{S}^{(i)} = (1-\eta)\mathbf{S}^{(i-1)} + \eta(\M^{(i-1)t}\M^{(i-1)})^{-1}\M^{(i-1)t}\D,
\end{equation}
\begin{equation}\label{eqMiter}
\M^{(i)} = (1-\eta)\M^{(i-1)} + \eta(\mathbf{S}^{(i)}\mathbf{S}^{(i)t})^{-1}\mathbf{S}^{(i)}\D^t,
\end{equation}
where $0<\eta\leq1$ is a step size. Intuitively, at each iteration we are computing the least square solution $\mathbf{S}^{(i)}$ to \refeq{eqMWD}, treating $\M^{(i-1)}$ as the truth, and vice versa. We keep iterating until the cost function decreases by less than $0.01\%$.

Due to incomplete sky coverage, there are two further tweaks to the above equations. Firstly, we do not directly compute the above equations in matrix form. Rather, we compute $\mathbf{S}^{(i)}$ column by column, with each column representing one frequency, and we modify $\M^{(i-1)}$ to exclude the pixels not covered at that frequency. Similarly for \refeq{eqMiter}, each column corresponds to a pixel, and we modify $\mathbf{S}^{(i)}$ to exclude frequencies that do not cover that pixel. Thus, at every iteration, the algorithm incorporates all pixel data in $\D$, regardless of the sky coverage at each frequency or the frequency coverage of each pixel. In addition, since the maps have different sky coverages, simply normalizing valid pixels in each map will over-weigh the maps covering low temperature out-of-plane regions compared to those that only cover the Galactic centre. Thus, we re-normalise the data matrix $\D$ at every iteration by dividing each map by the norm (root-sum-square) of its best-fitting map $\M^{(i)}\mathbf{S}^{(i)}$.

%%%%%%%%%%%%%%%%%%%%%%%%%%%%%
%%%%%%%%%%%%%%%%%%%%%%%%%%%%%
\subsection{Incorporating More General Data Formats}\label{subsecinterferometric}
\refeq{eqMWD} assumes that all data sets contained in $\D$ are sky maps with the same pixelization and angular resolution, which is what we focus on in this work. However, this comes with the drawback of forcing us to work with the lowest common resolution for all maps, thus discarding the high frequency information in higher quality maps. Furthermore, many low-frequency interferometers such as \textit{MWA} \citep{MWA} and \textit{PAPER} \citep{PAPER, PAPERpspec, PAPERpspec2} have produced high quality data, but because they sample sparsely in the Fourier domain, it is challenging to reduce those measurements into pixelised sky maps. Fortunately, we can generalise our algorithm and allow all data sets that are linearly related to the pixelised sky maps. The data set at frequency index $f$ can be described by
\begin{equation}\label{eqMAWD}
D_{if} = \sum_{jn}B_{ij}^f M_{jn} S_{nf},
\end{equation}
which is a generalised version of \refeq{eqMWD}. At any given frequency index $f$, $\B^f$ is a known matrix describing the linear relationship between the pixelised sky map and the data set. For pixelised sky-maps, $\B^f$ is the identity matrix and \refeq{eqMAWD} reduces to \refeq{eqMWD}. For low resolution data sets, $\B^f$ contains the point spread function for all pixels in the sky, which is typically the antenna beam pattern. For drift-scanning interferometer data sets, $\B^f$ contains the primary beam pattern as well as the rotating fringe patterns for all baselines, and $D_{fi}$ at $f$ is a flattened list of visibilities over both time and baselines. Given $\B$ and $\D$, one can iteratively solve for $\mathbf{S}$ and $\M$. It is worth noting that while including the $\B$-matrix does not significantly increase the amount of computations for obtaining $\mathbf{S}$ at each iteration, it makes the computation of $\M$ much more demanding. This is because one can no longer use \refeq{eqMiter} to compute $\M$ pixel by pixel, as the $\B$ mixes pixels in the sky. One now needs to operate matrices with $n_\text{pix}\times n_\text{c}$ rows rather than $n_\text{c}$ rows as in \refeq{eqMiter}. For a HEALPix map \citep{HEALPIX} with $n_{\rm side}=64$ corresponding to pixel size of $1^\circ$ and 6 principal components, the matrix size is about $3\times 10^5$ on each side, which demands significant computing resources beyond the scope of this work.

%%%%%%%%%%%%%%%%%%%%%%%%%%%%%
%%%%%%%%%%%%%%%%%%%%%%%%%%%%%
%%%%%%%%%%%%%%%%%%%%%%%%%%%%%

\begin{figure*}
	\centerline{\includegraphics[width=\textwidth]{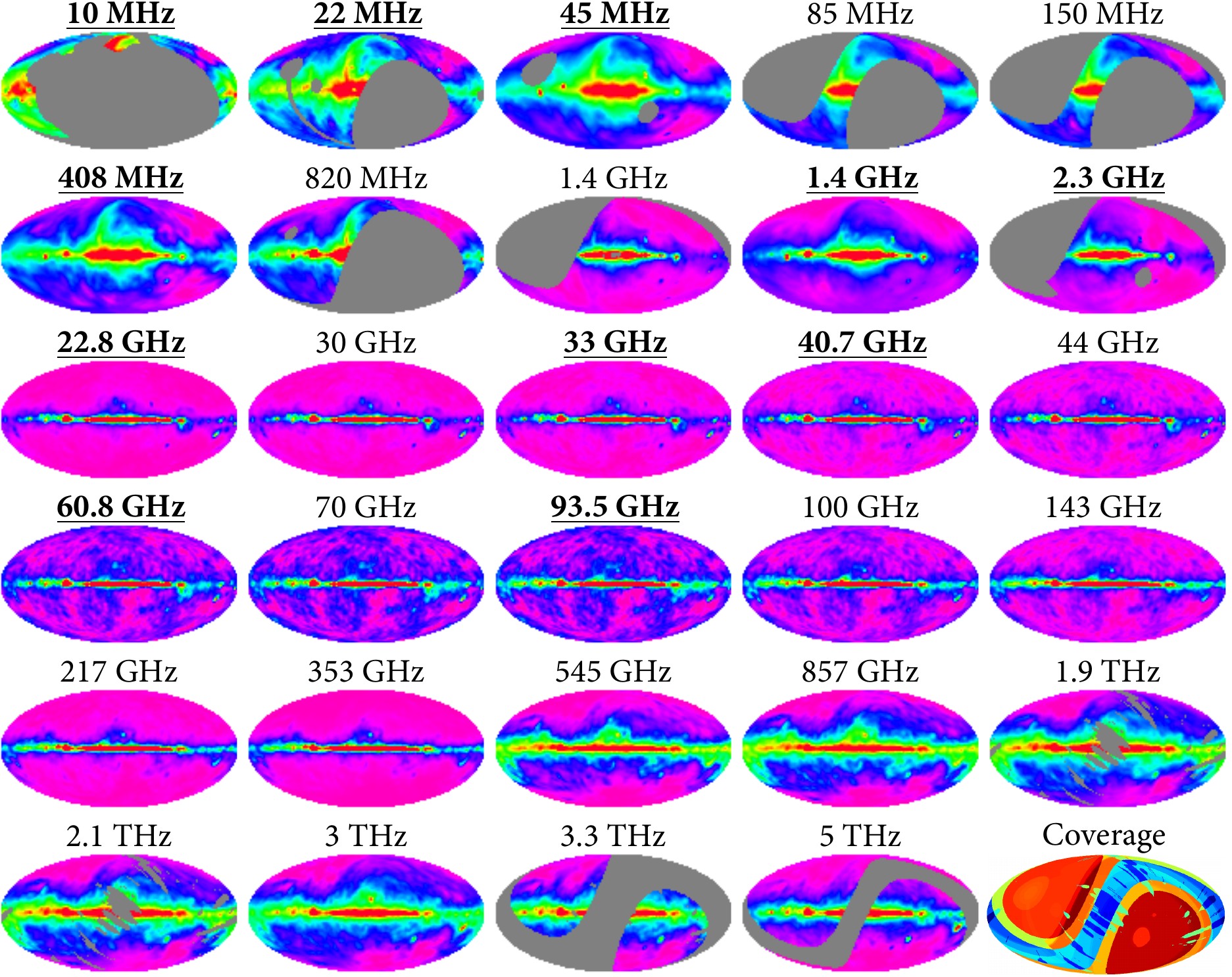}}
	\caption[29 sky maps used in this work from 10\,MHz to 5\,THz.]{
		29 sky maps used in this work from 10\,MHz to 5\,THz, plotted in Galactic coordinate under Mollweide projection centred at the Galactic centre, on arcsinh scales, where the constant for each arcsinh is set to the overall amplitude of each map, as shown in \reffig{fignorm}. The color scale follows the rainbow order, with red being the highest, purple the lowest, and gray signifying no data. The 11 bold and underscored frequencies are those included in the original GSM. The last panel shows the 120 different frequency coverage regions, each represented by a different colour (the progression of colour implies no particular ordering), and none of which contains all 29 frequencies. 
		\label{figallmaps}
	}
\end{figure*}

\begin{figure}
	\centerline{\includegraphics[width=.48\textwidth]{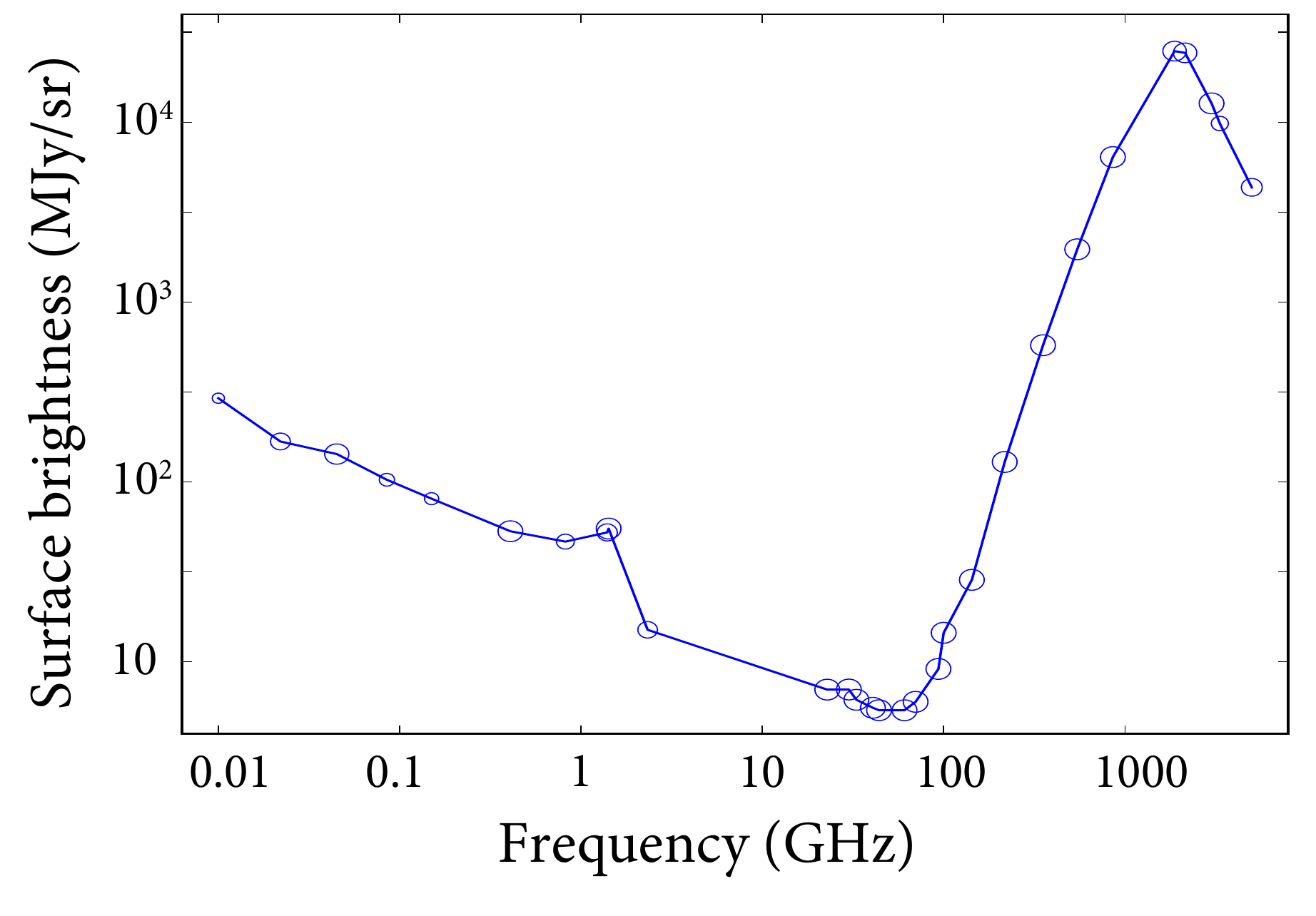}}
	\caption[The overall amplitudes of the 29 sky maps.]{
		The overall amplitudes of the 29 sky maps included in this work. The sizes of the circles represent the sky coverage of each map. The amplitudes can be thought of as the root-mean-square value of the input maps, with some iterative adjustments explained in \refsec{subseciterativegsm2}. 
		\label{fignorm}
	}
\end{figure}

\section{Sky Survey Data Sets}\label{secsurvey}
Table~\ref{taballdata} and \reffig{figallmaps} lists all sky maps we use in this work, ranging from 10\,MHz to 5\,THz. We have included all sky maps in this frequency range that have angular resolutions better than $5^\circ$ and sky coverage larger than 20\% of the full sky. All data sets are publicly available online\footnote{MPIfR: \url{www3.mpifr-bonn.mpg.de/survey.html},\\ LAMBDA: \url{lambda.gsfc.nasa.gov/product/},\\ \textit{Planck}: \url{pla.esac.esa.int/pla}.}, except those at 45\,MHz and 2.33\,GHz, which are available on request from the respective data owners. We manually mask out the ecliptic plane in the 3.33\,THz \textit{AKARI} data and the 5\,THz \textit{IRIS} data, and exclude \textit{IRIS} data above 5\,THz, in order to remove zodiacal contamination. Unlike the original GSM, which used physical models to pre-remove the CMB anisotropy from the \textit{WMAP} and \textit{Planck} maps, we choose to let the data speak for themselves and not pre-remove the CMB anisotropy. These 29 maps form regions with 120 different combinations of map overlap, with no common region between all maps (see the bottom right plot in \reffig{figallmaps}), so the improved GSM-making method is necessary to combine all these data sets.

\begin{table*}
	
	\centering

	\begin{tabular}{  m{0.3\columnwidth}  m{0.2\columnwidth}  m{0.1\columnwidth}  m{0.2\columnwidth}  m{0.9\columnwidth}  }
		%		\begin{tabular}{  m{3cm}  m{1.2cm}  m{0.8cm}  m{2cm}  m{1.1cm}  }
		\hline
		\hline
		\vskip0.65mm
		Project/Instrument & $\nu(\text{GHz})$ & Area & Resolution & Reference(s)\\
		\hline
		\vskip0.65mm
		DRAO, CAN & 0.01 & N & $2.6\times 1.9^{\circ}$& \citet{Caswell1976DRAO}\\ 
		\vskip0.65mm
		DRAO, CAN & 0.022 & N & $1.1\times 1.7^{\circ}$& \citet{Roger1999DRAO}\\ 
		%		\vskip0.65mm
		%		MRAO & 0.045 & S & $4.6\times 2.4^{\circ}$& \citep{Alvarez1997MRAO}\\ 
		\vskip0.65mm
		MRAO+JMUAR & 0.045 & F & $\sim3.6^{\circ}$& \citet{Alvarez1997MRAO, Maeda1999JMUAR}\\ 
		\vskip0.65mm
		Parkes & 0.085 & E & $3.8\times 3.5^{\circ}$& \citet{Landecker1970Parkes}\\ 
		\vskip0.65mm
		Parkes & 0.15 & E & $2.2^{\circ}$& \citet{Landecker1970Parkes}\\ 
		\vskip0.65mm
		GER, AUS, ENG & 0.408 & F& $56'$ & \citet{Haslam1981, Haslam1982, Remazeilles2015Haslam}\\ 
		\vskip0.65mm
		Dwingeloo, NLD & 0.82 & N & $1.2^{\circ}$& \citet{Berkhuijsen1971}\\ 
		\vskip0.65mm
		\textit{CHIPASS} & 1.39 & S & $14.4'$ &\citet{Calabretta2014CHIPASS}\\ 
		\vskip0.65mm
		Stokert, Villa Elisa& 1.42 & F & $36'$& \citet{Reich1982Stockert, Reich1986Stockert, Reich2001Elisa}\\ 
		%\vskip0.65mm
		%S-PASS & 2.3 & S & $9'$& ???\\
		\vskip0.65mm
		Rhodes/HartRAO & 2.33 & S & $20'$& \citet{Jonas1998Rhodes}\\ 
		\vskip0.65mm
		%Stockert & 2.72 & P & $4.3'$& \citep{Reif1984stockert11cm2, Reif1990stockert11cm3, Reif1990stockert11cm4}\\ 
		%\vskip0.65mm
		\textit{WMAP} & 22.8 & F & $49'$& \citet{WMAP5year}\\ 
		\vskip0.65mm
		\textit{Planck} & 28.4 &F& $32'$ & \citet{planck2015}\\ 
		\vskip0.65mm
		\textit{WMAP} & 33.0 & F & $37'$& \citet{WMAP5year}\\ 
		\vskip0.65mm
		\textit{WMAP} & 40.6 & F & $29'$& \citet{WMAP5year}\\ 
		\vskip0.65mm
		\textit{Planck} & 44.1 &F& $24'$ & \citet{planck2015}\\ 
		\vskip0.65mm
		\textit{WMAP} & 60.8 & F & $20'$& \citet{WMAP5year}\\ 
		\vskip0.65mm
		\textit{Planck} & 70.4 &F& $14'$ & \citet{planck2015}\\ 
		\vskip0.65mm
		\textit{WMAP} & 93.5 & F & $13'$& \citet{WMAP5year}\\  
		\vskip0.65mm
		\textit{Planck} & 100 &F& $10'$ & \citet{planck2015}\\  
		\vskip0.65mm
		\textit{Planck} & 143 &F& $7'$ & \citet{planck2015}\\  
		\vskip0.65mm
		\textit{Planck} & 217 &F& $5'$ & \citet{planck2015}\\  
		\vskip0.65mm
		\textit{Planck} & 353 &F& $5'$ & \citet{planck2015}\\  
		\vskip0.65mm
		\textit{Planck} & 545 &F& $5'$ & \citet{planck2015}\\  
		\vskip0.65mm
		\textit{Planck} & 857 &F& $5'$ & \citet{planck2015}\\ 
		\vskip0.65mm
		\textit{AKARI} & 1875 & P& $1.5'$ & \citet{Doi2015AKARI}\\ 
		\vskip0.65mm
		\textit{AKARI} & 2143 & P& $1.5'$ & \citet{Doi2015AKARI}\\ 
		\vskip0.65mm
		\textit{IRAS} (\textit{IRIS}) & 3000 & F& $4.3'$ & \citet{Miville2005IRIS}\\
		\vskip0.65mm
		\textit{AKARI} & 3333 & P& $1.5'$ & \citet{Doi2015AKARI}\\ 
		\vskip0.65mm
		\textit{IRAS} (\textit{IRIS}) & 5000 & P& $4'$ & \citet{Miville2005IRIS}\\
		\hline
	\end{tabular}
	\caption[List of sky maps we use in our multi-frequency modeling.]{
		List of sky maps we use in our multi-frequency modeling. F: full sky; S: southern sky; N: northern sky; E: equatorial plane; P: partial sky. \textit{CHIPASS} at 1.39\,GHz has a bandwidth of 64\,MHz, so its frequency largely overlaps with the Stokert+Villa Elisa map at 1.42\,GHz.
		\label{taballdata}
	}
\end{table*}

Due to the large frequency range covered by these sky maps and the variety of instruments involved in making them, there is considerable variation in their characteristics. For example, some instruments have variable beam sizes, while others have significant sidelobes. For this work, we approximate all beams as Gaussian, whose full-width-half-maximum size ($\theta_\text{FWHM}$) equal the resolution quoted in Table~\ref{taballdata}. We smooth all maps to $5^\circ$ by applying Gaussian smoothing kernels of sizes $\sqrt{{5^\circ}^2-\theta_\text{FWHM}^2}$, and remove another $3^\circ$ from edge areas in incomplete maps. All map are pixelised onto a HEALPix grid with $n_{\rm side}=64$.

Additionally, each instrument faces unique calibration challenges, which causes them to differ in error properties. While it is possible to quantify various error properties in the form of error covariance matrices and insert them into \refeq{eqWiter} and \refeq{eqMiter}, we leave such effort to future work.

In terms of unit, the sky maps come in three different units. The maps below 20\,GHz are in Rayleigh-Jeans temperature, $\text{K}_\text{RJ}$, below 500\,GHz and above 20\,GHz in CMB temperature, $\text{K}_\text{CMB}$, and above 500\,GHz in MJy/sr. We choose to convert all maps into to MJy/sr using the following conversion equations:
\begin{equation}
\frac{1 \text{K}_\text{RJ}}{1 \text{MJy/sr}} = 2 \times 10^{20} \frac{k_\text{B}}{\lambda^2},
\end{equation}
\begin{equation}
%\frac{1 \text{K}_\text{CMB}}{1 \text{MJy/sr}} = 2 \times 10^{20} \frac{k_\text{B}}{\lambda^2}\left(\frac{h\nu}{k_\text{B}T_\text{CMB}}\right)^2\frac{e^{\frac{h\nu}{k_\text{B}T_\text{CMB}}}}{\left( e^{\frac{h\nu}{k_\text{B}T_\text{CMB}}}-1\right)^2},
\frac{1 \text{K}_\text{CMB}}{1 \text{K}_\text{RJ}} = \left(\frac{h\nu}{k_\text{B}T_\text{CMB}}\right)^2\frac{e^{\frac{h\nu}{k_\text{B}T_\text{CMB}}}}{\left( e^{\frac{h\nu}{k_\text{B}T_\text{CMB}}}-1\right)^2},
\end{equation}
where $\nu$ is the frequency at the centre of the frequency band of the map, $\lambda = c/\nu$ is the wavelength, $k_\text{B}$ is Boltzmann's constant, $h$ is Planck's constant, and $T_\text{CMB}$ is the CMB temperature. The RMS amplitude of all the maps after unit conversion are shown in \reffig{fignorm}.

%%%%%%%%%%%%%%%%%%%%%%%%%%%%%
%%%%%%%%%%%%%%%%%%%%%%%%%%%%%
%%%%%%%%%%%%%%%%%%%%%%%%%%%%%
\section{Results I: the Improved GSM}\label{secresult}

%%%%%%%%%%%%%%%%%%%%%%%%%%%%%
%%%%%%%%%%%%%%%%%%%%%%%%%%%%%
\subsection{Orthogonal Components Result}\label{subsecorthogonal}
\begin{figure}
	\centerline{\includegraphics[width=.9\columnwidth]{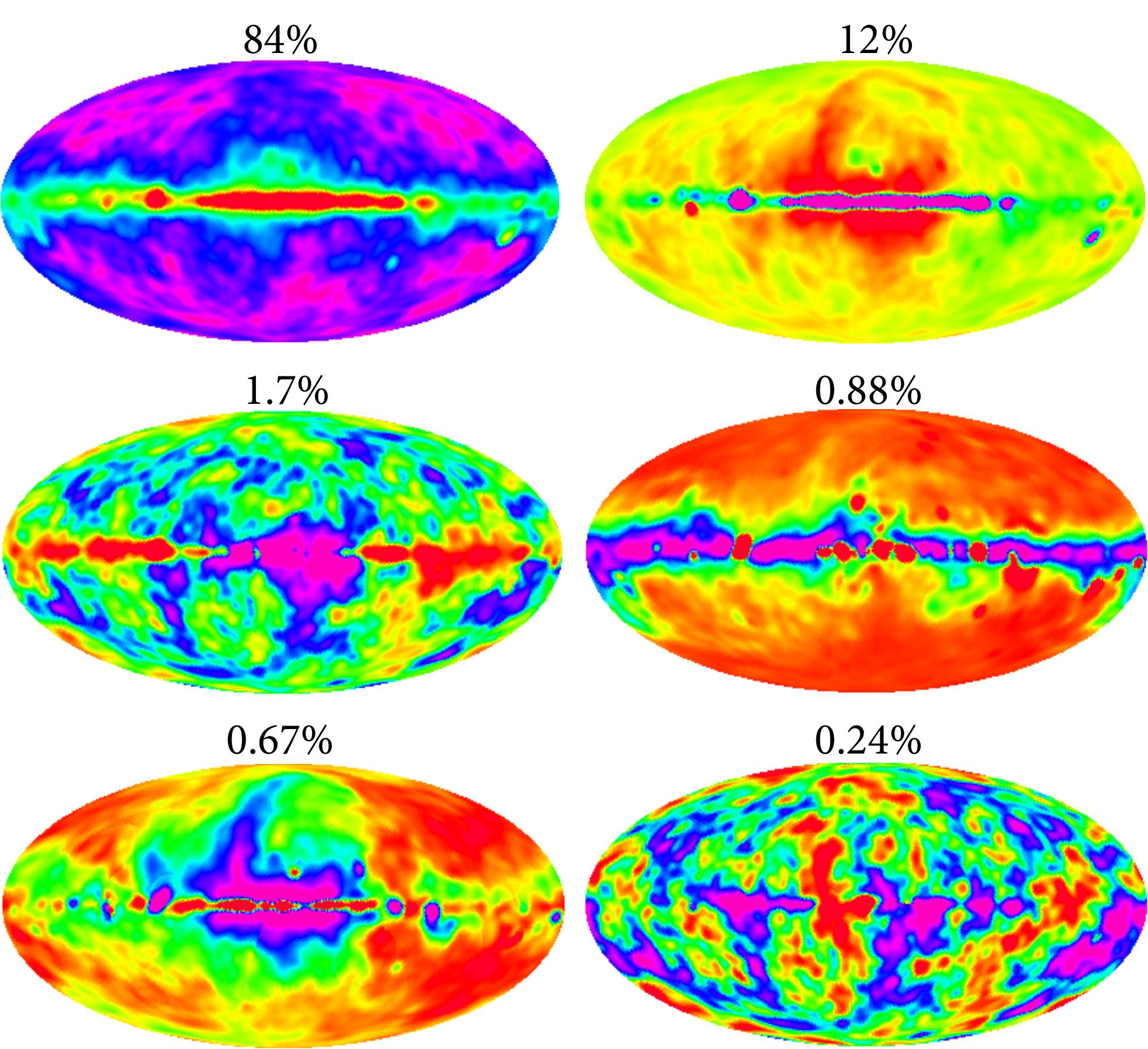}}
	\centerline{\includegraphics[width=.45\textwidth]{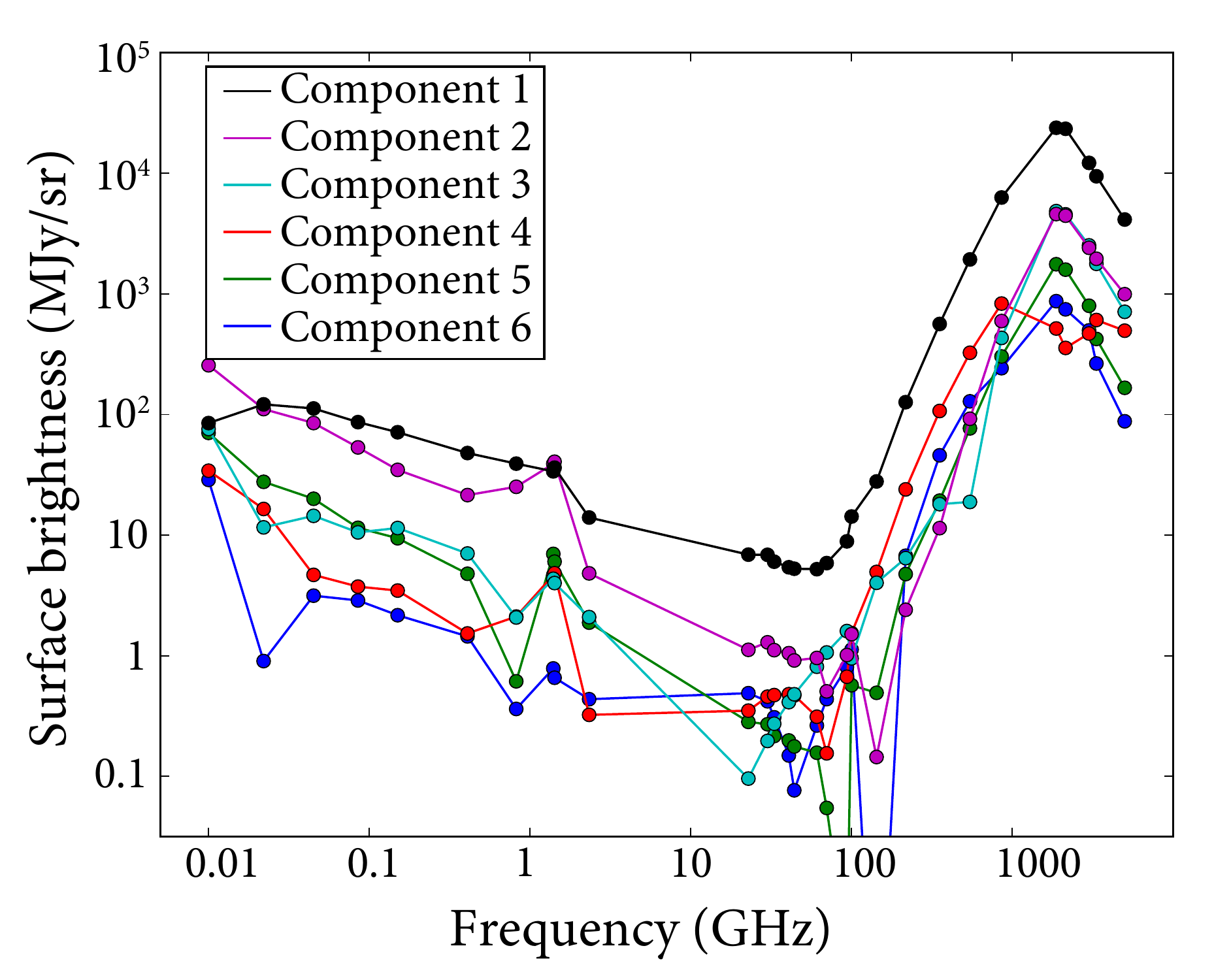}}
	\caption[The 6 orthogonal components and their spectra.]{
		\textit{Top:} the 6 orthogonal components. \textit{Bottom:} their spectra. The maps are plotted on arcsinh scales where the constant in arcsinh is set to the median of pixel amplitudes. The percentage on each component represents the fraction of all variations among the 29 input maps explained by that component. Here the spectra are not directly $\mathbf{S}$, but the rows of $\mathbf{S}$ multiplied by the overall amplitudes of the input maps shown in \reffig{fignorm}. Note that the spectra here have no physical meanings attached to them, which we explore in \refsec{secrecombine}.
		\label{figcomponentsorth}
	}
\end{figure}

We apply the algorithms described in \refsec{secalgorithm} on maps described in \refsec{secsurvey}. To obtain the initial 5\% common coverage, maps at 10\,MHz, 85\,MHz, 150\,MHz, and 5\,THz are temporarily excluded. In the iteration process, it takes 12 iterations with step size $\eta=1$ to converge. We also tried a smaller $\eta=0.2$, and obtained the same result with 4 times more iterations. After the first convergence, we mask out the 1\% of the pixels with the highest fitting residual in order to eliminate point sources, and re-iterate for another 8 iterations to arrive at the final convergence. The new GSM with six components is shown in \reffig{figcomponentsorth}. We choose the number of components to be six based on the predictive accuracy of the resulting GSM, which we discuss in more detail in \refsec{subsecerrorgsm2}. We find that the increase in predictive accuracy from using more than six components is negligible compared to the fitting errors. We plan to perform a more rigorous model selection to determine the optimal number of components in a future work that will incorporate uncertainty information from each input map.

%%%%%%%%%%%%%%%%%%%%%%%%%%%%%
%%%%%%%%%%%%%%%%%%%%%%%%%%%%%
\subsection{Error Analysis}\label{subsecerrorgsm2}
\begin{figure*}
	\centerline{\includegraphics[width=\textwidth]{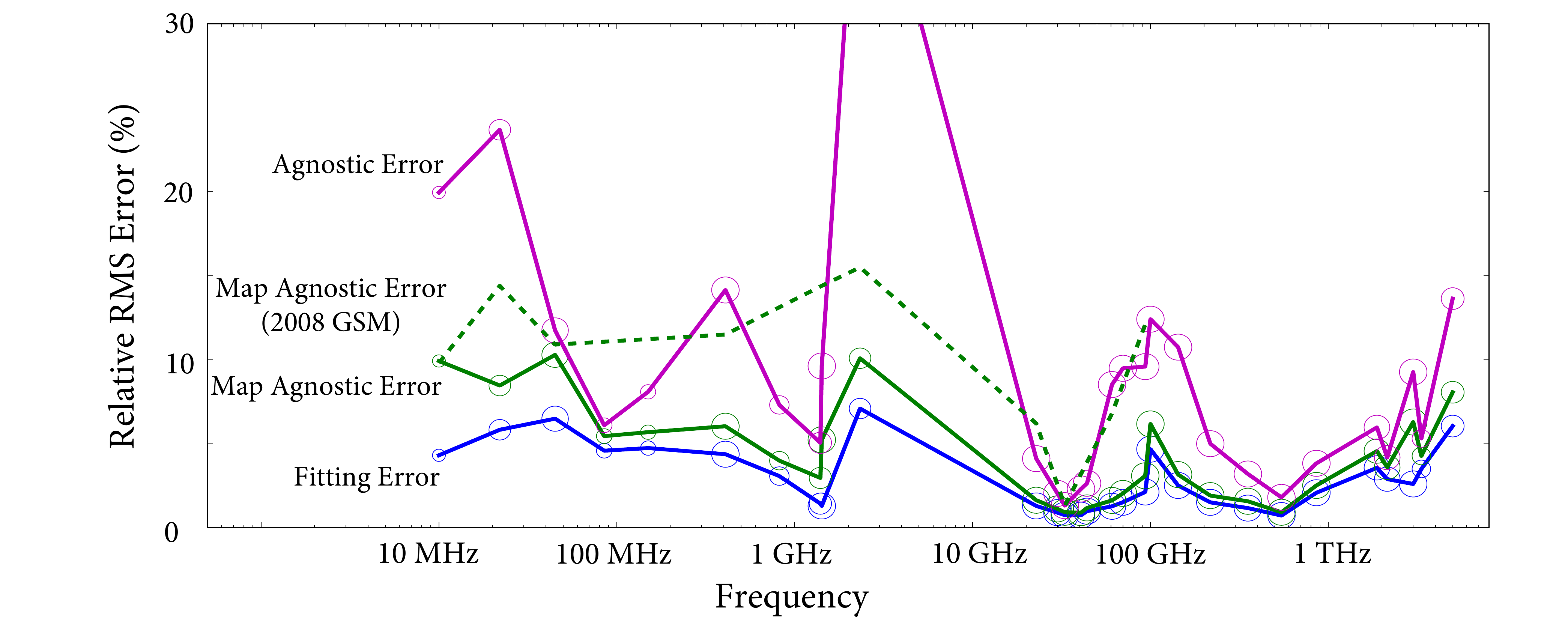}}
	\caption[Three different RMS error percentage estimations for our GSM.]{
		Three different RMS error percentage estimations for our GSM, compared to the error quoted with the original GSM. Agnostic error is an overly pessimistic upper bound to the predictive accuracy of the GSM, whereas the map agnostic error is the lower bound. Fitting error is the fitting residual in each map not explained by the GSM. The sizes of the circles represent the sky coverage of each map. The improved GSM has less error compared to the original GSM across the entire frequency range.
		\label{figgsm2error}
	}
\end{figure*}

To assess the predictive power of the GSM, we compute three types of Root Mean Square (RMS) errors at each frequency. The first is the fitting residual, which is the difference between the complete GSM model and the input sky maps. This fitting error is typically 1\% for CMB frequencies between 20 and 100\,GHz, and around 5\% everywhere else, as shown in blue in \reffig{figgsm2error}. This is a combination of the measurement errors in the input maps and weak emission mechanisms comparable to such errors, which we are unable to capture. The fitting error represent the lower bound of the predictive accuracy of our GSM.

The second type of error is the map ``agnostic error'', where at a given frequency, we use the same normalised spectra but do not use the input map at that frequency to calculate the principal component maps. We calculate the RMS of the difference between the resulting model and the actual sky map. As shown in green in \reffig{figgsm2error}, this error is typically 1.5 to 2 times larger than the fitting error. Among the three errors discussed in this section, this error is the most probable estimate of the predictive accuracy of the GSM, and is also used in the original GSM work.

The third type of error is a conservative upper-bound to the predictive accuracy of the GSM. Here we pretend to have zero knowledge of each map and exclude it from the algorithm from the outset, and produce a new GSM that is agnostic of that map. We then compare the excluded map with the predicted map using the new agnostic GSM, and calculate the RMS of the difference after an overall renormalization. The renormalization is typically less than 15\%. This is an overly pessimistic estimate of the predictive power of the GSM, because we are not in the regime of an over-abundance of sky maps, especially at frequencies below 20\,GHz. The only two full-sky maps below CMB (at 408\,MHz and 1.4\,GHz) serve as ``cornerstones'' of the GSM, so the agnostic error caused by removing them completely (and thus crippling the GSM) is not a good estimate of the predictive power of the complete GSM at an unexplored frequency such as 300\,MHz.

The peaks in the agnostic error curves that rise above 10\% fall into three categories. The first category includes those near spectral lines, where the high errors are caused by having a spike at one end of the interpolation. One such example is the strongest peak at 2.3\,GHz, whose error mostly comes from interpolating between the peak at 1.4\,GHz and the lowest \textit{WMAP} frequency. Another example is the peak around 100\,GHz, which is contaminated by the 115\,GHz CO line. The second category includes those at the edge of our frequency range, namely 10\,MHz and 5\,THz. They have high errors due to extrapolation. The third category includes those that interpolate between incomplete and typically non-overlapping maps. One example is the 408\,MHz Haslam map, which is interpolated between the 150\,MHz equatorial map and the 820\,MHz northern sky map. Another example is the 3\,THz \textit{IRIS} map that is interpolated between two \textit{AKARI} maps with poor equatorial coverage.

In summary, for a map predicted by the GSM, we expect the RMS accuracy to be above the fitting error, and below the completely agnostic error. This is typically between 5\% and 15\% for most frequencies, and for frequencies near 50\,GHz and 500,\GHz, the accuracy can be as good as 2\%. In addition, we expect an overall amplitude offset of less than 15\%. In addition to a wider and denser frequency coverage, the improved GSM is seen to be consistently more accurate, by up to a factor of 2, than the original GSM across the entire frequency range.

%%%%%%%%%%%%%%%%%%%%%%%%%%%%%
%%%%%%%%%%%%%%%%%%%%%%%%%%%%%
%%%%%%%%%%%%%%%%%%%%%%%%%%%%%
\section{Results II: Recombine the GSM to Extract Physical Components}\label{secrecombine}

%%%%%%%%%%%%%%%%%%%%%%%%%%%%%
%%%%%%%%%%%%%%%%%%%%%%%%%%%%%
\subsection{Blind Component Separation}\label{subsecphysical}
Since the component maps we obtain (\reffig{figcomponentsorth}) are mutually orthogonal by construction, they are linear combinations of underlying emission mechanisms such as synchrotron and thermal dust. As discussed in \refsec{subsecframework}, we have the freedom to apply any $6\times6$ invertible matrix $\mPsi$ and its inverse to the matrices $\mathbf{S}$ and $\M$ without changing our model, $\M\mathbf{S}$. Applying $\mPsi$ will not change any linear interpolation results either. In this section we discuss our two-step automatic algorithm to determine $\mPsi$ and blindly extract the physical contributions. This algorithm focuses on manipulating the normalised spectrum matrix $\mathbf{S}$. The only two pieces of information that we use in order to determine $\mPsi$ are: 1) the absence of monopole in CMB; and 2) compact frequency support for each normalised spectrum, meaning that the each component dominates a limited frequency range. Our algorithm is thus blind to any existing models for known emission mechanisms. 

We start with a $\mPsi$ that equals the identity matrix. The first step tries to search for the smallest frequency range outside which there is only five components, meaning that the sixth component must be limited to said frequency range. To do this, we enumerate all possible frequency ranges, and for each frequency range, we remove the columns in $\mathbf{S}$ that correspond to that range to form $\mathbf{S}^*$. We then calculate the eigenvalues and eigenvectors for $\mathbf{S}^*\mathbf{S}^{*t}$. For the smallest frequency range whose smallest eigenmode explains less than 1\% of all variations in $\mathbf{S}^*$,
\begin{equation}
\frac{\lambda_\text{min}}{\text{tr}(\mathbf{S}^*\mathbf{S}^{*t})} < 1\%,
\end{equation}
we multiply $\mPsi$ by its eigenvectors $\C^*$. This way, the last row of $\mPsi\mathbf{S}$, which corresponds to the smallest eigenvalue, is our first separated component. If there is no frequency range that satisfies this criteria, we keep doubling the 1\% threshold until one valid range appears. In practice, the synchrotron component is separated first with 2\% threshold. We then repeat the above procedure, and separate the CMB component at 4\%. Lastly, the ``HI'' component is separated at 8\%. The first step ends since we cap the threshold at 10\%.

Before we proceed to the second step, we attempt to clean the foreground contamination in the CMB component found in the first step. To do this we find the $\mPsi$ that minimises the power in the CMB component map. This is the only procedure in the entire algorithm where we are working with the component map rather than normalised spectra, also the only procedure that use physical knowledge, namely the lack of monopole in the CMB.

\begin{figure}
	\centerline{\includegraphics[width=.9\columnwidth]{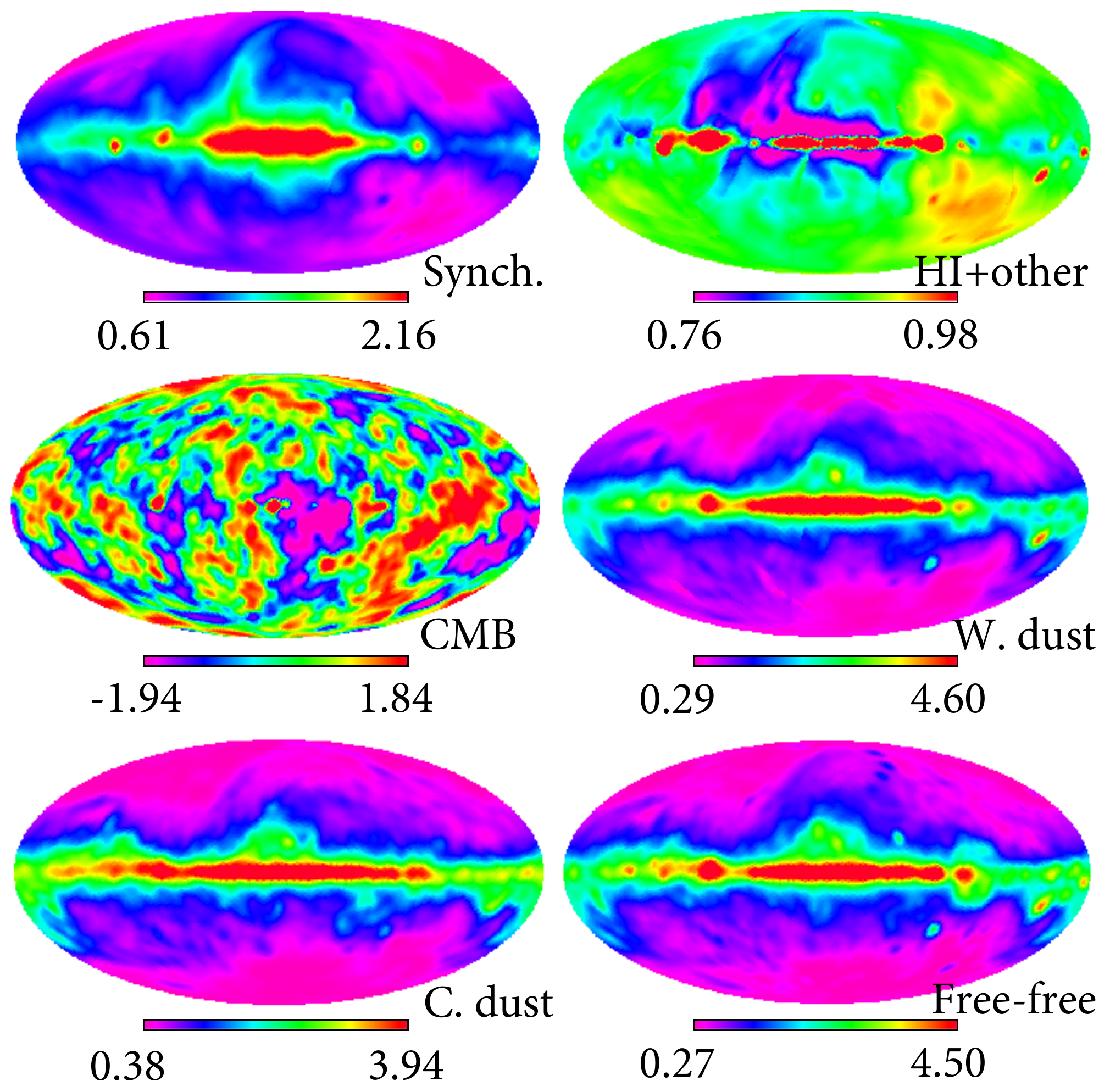}}
	\centerline{\includegraphics[width=.9\columnwidth]{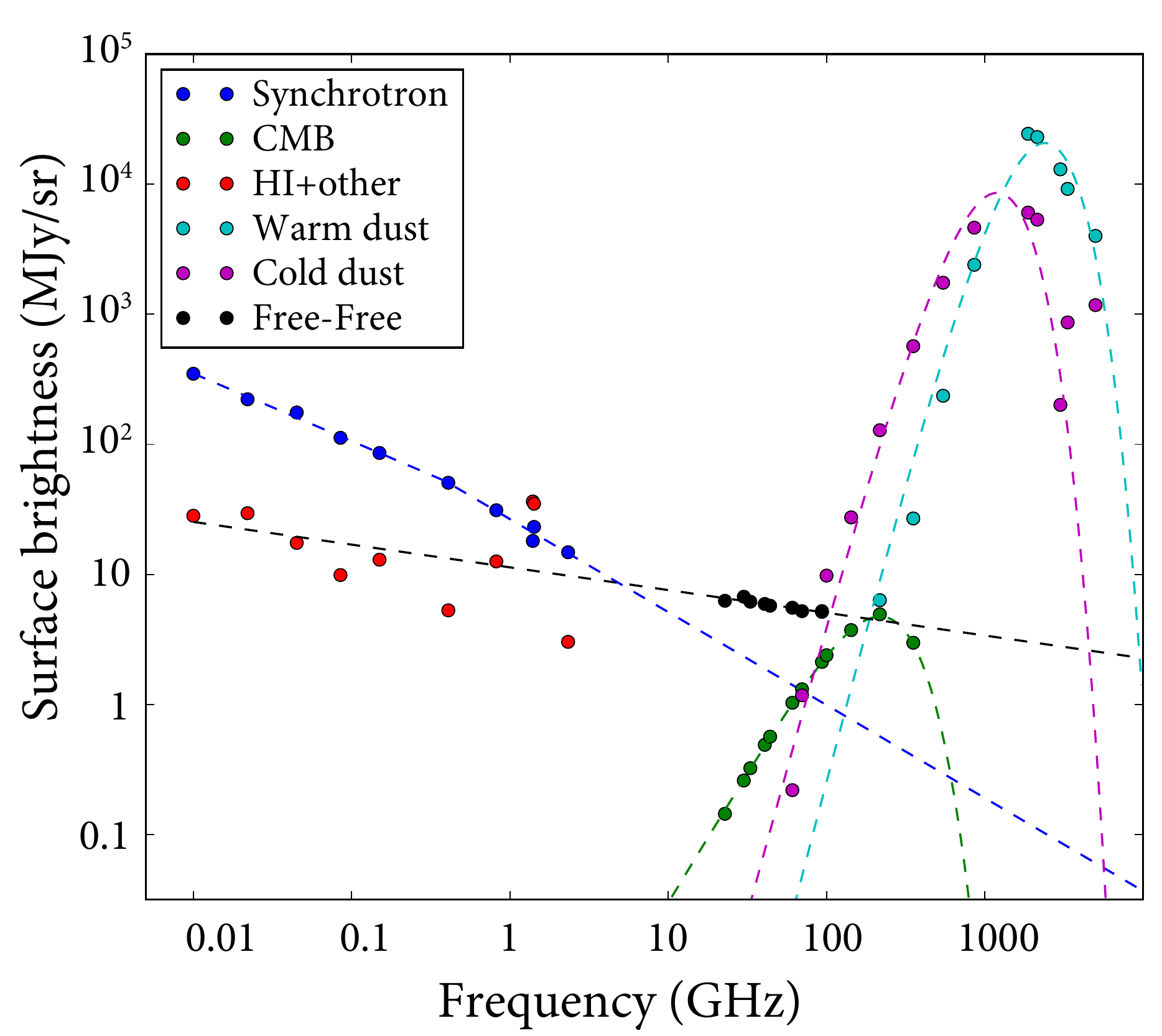}}
	\caption[The 6 recombined components that roughly correspond to various physical mechanisms.]{
		\textit{Top:} six component maps from blind component separation. \textit{Bottom:} six component spectra (solid points) and their best fits (dashed lines). For plotting purpose, each component map is normalised to have a median of 1 and plotted on arcsinh scales, where the colour range corresponds to 2 and 98 percentiles in each map. In the bottom panel, we only plot the two most dominant component spectra at each frequency to reduce visual clutter. The best-fitting parameters are listed in Table~\ref{tabpars}.
		\label{figcomponentsphys}
	}
\end{figure}

The automatic procedure mentioned above is not able to separate the last three modes, which suggests that they cover largely overlapping frequency ranges. For our second step, we demand that the last three components each dominate one of the three frequency ranges: 10\,GHz to 100\,GHz, 100\,GHz to 1\,THz, and above 1\,THz. To do this, we find the $\mPsi$ that minimises the normalised spectrum of each component outside its designated range. The results we obtain are shown in solid dots in \reffig{figcomponentsphys}.

%%%%%%%%%%%%%%%%%%%%%%%%%%%%%
%%%%%%%%%%%%%%%%%%%%%%%%%%%%%
\subsection{Fitting the Blind Components}\label{subsecphysicalspectra}
As shown in \reffig{figcomponentsphys}, five of the six components obtained in our blind separation have spectra that closely resemble those of known physical processes: synchrotron, free-free, CMB, and two thermal dust species at different temperatures. In this section, we take the known spectrum models of these mechanisms, such as a power law for synchrotron, and find the best parameters for them that fit our blind spectra. We first compute six spectra by taking the new normalised spectrum matrix $\mPsi\mathbf{S}$ and multiplying back the normalization (shown in \reffig{fignorm}). We then fit each spectrum using its corresponding physical model: synchrotron and free-free as power laws with spectral indices $\beta$, thermal dust as blackbody spectra multiplied by power laws with temperatures $T$ and spectral indices $\beta$, and CMB anisotropy as the first order Taylor expansion of the blackbody spectrum with temperature $T$. Note that we do not change our blind spectra based on these models.

The fits are performed in log(frequency)-log(surface brightness) space, and input variances are empirically estimated using the root mean square of best-fit residuals. We limit the fitting range for each component to the frequency range it dominates, as shown in \reffig{figcomponentsphys}. For power law fits, we perform a standard linear regression to obtain the best fits and error bars. For the dust and CMB components, we first numerically search for the best fit, and then perform Monte-Carlo on the input spectra using the empirical variances to estimate the error bars. Although our component separation is blind to any of these model spectra, the best-fitting parameters for all five of our blind spectra are seen to agree very well with existing literature values to within $2\sigma$, as shown in Table~\ref{tabpars}.
%For all the non-CMB models, the error bars on our best-fitting parameters are comparable, if not better, than the current best values in the literature. In addition to the spectra, their component maps also agree well with existing spatial templates, which we discuss in the next section.

\begin{table*}
	\begin{tabular}{  m{0.1\textwidth}  m{0.23\textwidth}  m{0.2\textwidth} m{0.15\textwidth}  m{0.22\textwidth}  }
		\hline
		\hline
		\vskip0.65mm
			Component & Parameter & Best-fit for blind spectra & Previous value & Ref.\\
		\hline
		\vskip0.65mm
			Synchrotron & spectral index $\beta$ below Haslam & $-2.519\pm0.018$ &$-2.5\pm0.1$&\citet{EDGESspectralindex}\\
		%\hline
		\vskip0.65mm
			& spectral index $\beta$ above Haslam & $-2.715\pm0.082$&$-2.695\pm0.120$&\citet{Lawson1987}\\
		\vskip0.65mm
		& &&&\citet{reichreich1988}\\
		\vskip0.65mm
		& &&&\citet{Davies1996}\\
		\vskip0.65mm
		& &&&\citet{platania2003}\\
		\hline
		\vskip0.65mm
			Free-free&spectral index $\beta$& $-2.175\pm0.032$& $-2.1\pm0.03$ &\citet{freefreeDickinson}\\
		\hline
		\vskip0.65mm
			CMB &temperature $T$  & $2.748\pm0.016$\,K&$2.72548\pm0.00057$\,K&\citet{cmbtemp}\\
		\hline
		\vskip0.65mm
			Warm dust &temperature  $T$ &$20.0\pm3.3$\,K&$15.95\pm0.25$\,K&\citet{dust1},\\
		%\hline
		\vskip0.65mm
			&spectral index $\beta$ &$2.78\pm0.70$&$2.76\pm0.06$&\citet{dust2}\\
		\hline
		\vskip0.65mm
			Cold dust &temperature $T$ &$10.4\pm1.2$\,K&$9.58\pm0.18$\,K&\citet{dust1},\\
		%\hline
		\vskip0.65mm
			&spectral index $\beta$ &$2.54\pm0.51$&$1.65\pm0.02$&\citet{dust2}\\
		\hline
	\end{tabular}
	\caption[List of model parameters for our blind components compared to existing literature values.]{
		List of model parameters for our blind components compared to previously published values. All parameters that fit our blind components agree with previously published values to within $2\sigma$. The error bar from \citet{platania2003} represents spatial variation rather than statistical uncertainty. The error bars for previously published dust model parameters are estimated using the difference in values between \citet{dust1} and \citet{dust2}.
		\label{tabpars}
	}
\end{table*}

For future versions of the GSM, a worthy goal would be to re-discover the Anomalous Microwave Emission (AME; \citealt{AME1, AME2, AME3, AME4, AME5, AME6, AME7, AME8, AME9, AME10, AME11, AME12}) component in addition to the five components mentioned above. Given the currently available data sets, the AME component is likely absorbed into both fitting error and other components such as thermal dusts. The absence of a stand-alone AME component can be attributed to insufficient signal-to-noise ratio, since AME peaks around 20 GHz, which is very close to the 3 to 20\,GHz frequency gap in existing data sets. Future GSM analysis with more data in the relevant frequency range will likely be able to separate the AME component.
%%%%%%%%%%%%%%%%%%%%%%%%%%%%%
%%%%%%%%%%%%%%%%%%%%%%%%%%%%%
\subsection{High Resolution GSM}\label{subsechighres}

\begin{figure*}
	\centerline{\includegraphics[width=\textwidth]{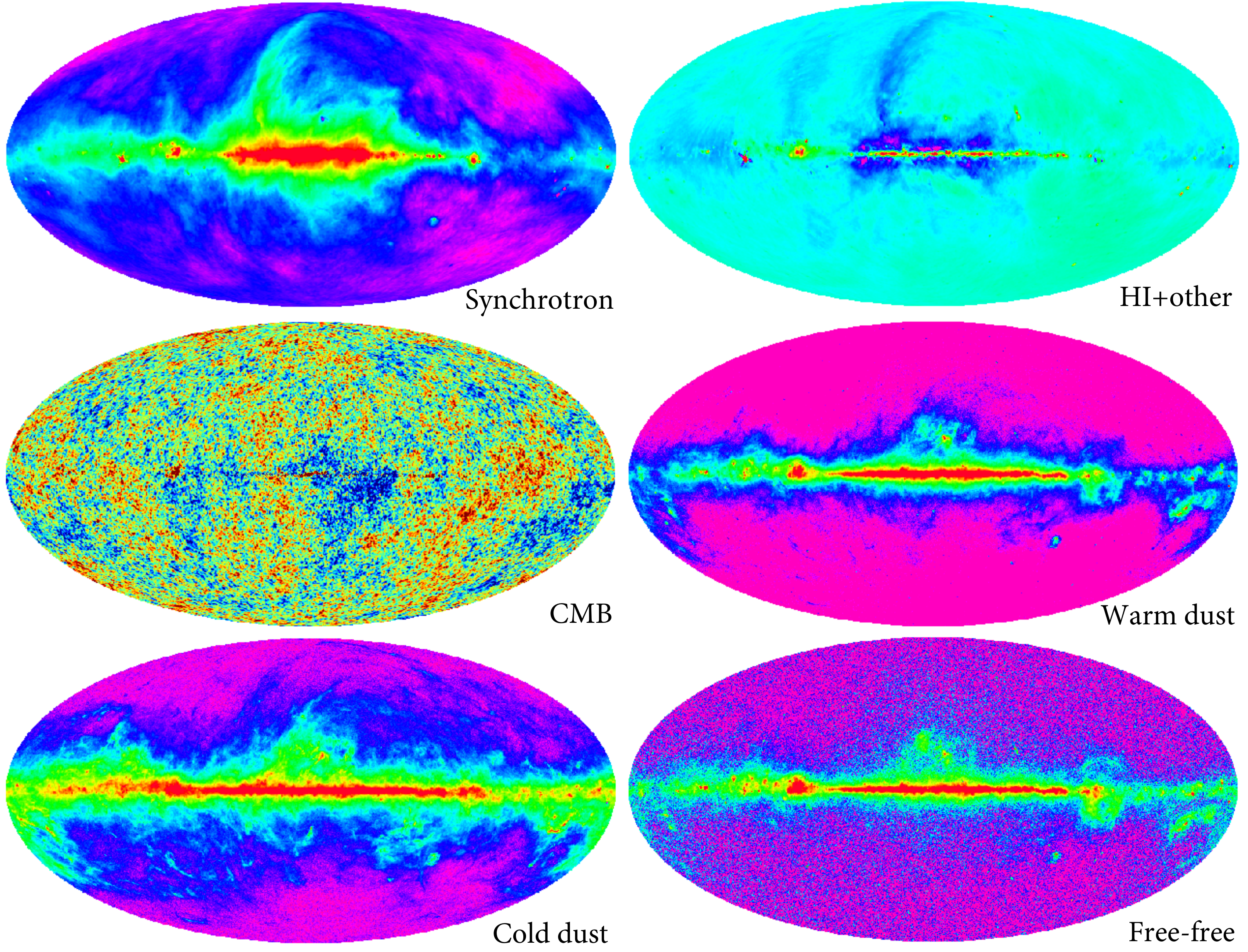}}
	\caption[High resolution version of the six component maps.]{
		High resolution version of the six component maps. The five non-CMB maps follow the same colour scales as those in \reffig{figcomponentsphys}, whereas the CMB component is plotted on a linear scale. The top two components have $56'$ angular resolution, and the other four $24'$.
		\label{fighighres}
	}
\end{figure*}

In addition to the $5^\circ$ resolution GSM we have discussed, we also produce a high resolution version using the same normalised spectra as the $5^\circ$ version. Thanks to the successful physical component separation as shown in \reffig{figcomponentsphys}, we are able to adopt two different resolutions throughout the frequency range. For frequencies above 40\,GHz, we can safely neglect the synchrotron component, because it is more than two orders of magnitude below the other four components at most of these frequencies, as shown in \reffig{figcomponentsphys}. We thus take those \textit{WMAP}, \textit{Planck}, \textit{AKARI}, and \textit{IRIS} maps whose resolution is better than 24 arcminutes, smooth them to 24 arcminutes, and solve for the CMB, the free-free, and the two dust component maps. Then, for frequencies below 10\,GHz, we take the 408\,MHz and 1.42\,GHz maps, smooth the latter to 56 arcminutes, smooth and remove the four component maps obtained in the high frequency band, and fit for the synchrotron and ``HI'' maps. Overall, the high resolution product has 56 arcminute angular resolution at frequencies below 10\,GHz, and 24 arcminute angular resolution above 10\,GHz, shown in \reffig{fighighres}. Note that while the high resolution component maps share the same spectra and large scale features as their low resolution counterparts, directly smoothing them to $5^\circ$ will not produce exactly the low resolution versions, because the high resolution component maps are calculated using only a subset of the 29 input maps.

Our blind high resolution component maps agree remarkably well with existing maps and spatial templates. Our synchrotron, CMB anisotropy, free-free, and cold dust maps share most of the features seen in their counterparts in both the \textit{WMAP} 9 year results and \textit{Planck} 2015 results (see e.g. Fig.~19 in \citet{wmap9} and Fig.~16 in \citealt{planck2015}). Free-free emission is known to closely trace H$\alpha$ emission \citep{Halpha}, and our free-free map shows all the key features visible in the composite H$\alpha$ map presented in \citet{Halpha}. Moreover, our cold dust map share all the key spatial features of the 94\,GHz dust map presented in \citet{dust1}.

%%%%%%%%%%%%%%%%%%%%%%%%%%%%%
%%%%%%%%%%%%%%%%%%%%%%%%%%%%%
\subsection{Comparison with \emph{Planck} and Other Foreground Models}\label{subsecphysdiscussion}
The blind spectra can be fitted very well in the frequency ranges they dominate, as shown in \reffig{figcomponentsphys} and Table~\ref{tabpars}. On the flip side, however, because our blind separation approach minimises the normalised spectrum of each component outside the frequency range it dominates, this approach cannot recover very well the parts of the spectrum outside its dominant range. One such example is the free-free spectrum, which is well fitted by a power law with spectral index of $-2.175\pm0.032$ near the CMB frequencies, but due to the minimization procedure, it largely disappears in the MHz range where synchrotron dominates. It is still contained in the GSM, but absorbed into the synchrotron and the HI+other components.

A similar problem is evident with our cold dust component. From Table~\ref{tabpars}, we see a significant difference in our best-fitting spectral index ($\beta = 2.54 \pm 0.51$) and accepted literature values (e.g., $\beta = 1.65 \pm 0.02$ from \citealt{dust2}). A visual inspection of Figure \ref{figcomponentsphys} suggests that this is a result of degeneracies with other components. One sees that from $100$ to $1000\,\textrm{GHz}$ (where cold dust is bright) is a rather ``crowded" region with many other components contributing non-negligibly to the sky emission. Moreover, the dominance of cold dust over these other components is weak at best. This makes our cold dust component particularly susceptible to leakage from components such as warm dust, driving our spectral index to a steeper value. We therefore see that our blind component extraction should be interpreted with some degree of caution when applied to frequency ranges where a wide range of foreground emission and mechanisms are at play. To some extent, the resulting degeneracies are captured by the large error bars associated with our dust parameters, but it is important to bear in mind that there may also be systematic errors arising from the intrinsic limitations of our method.

The sixth component, HI+other, is likely a combination of multiple mechanisms and systematic effects. Its strong peak near 1.4\,GHz suggests local 21\,cm emission as the dominant mechanism. In its component map, we notice two interesting features. There is a blue (low temperature) region whose shape resembles the synchrotron map, so this might indicate another weak synchrotron component with a different spectral index. We also notice a dipole component that is aligned with the equatorial poles, with striping artifacts especially visible near the northern celestial pole, which suggests contribution from scanning systematics in the low frequency maps.

To further examine our physical foreground components, we compare our data products with those derived from \emph{Planck} data. The Planck collaboration proposed a global representation of the multi-component sky over nine frequency bands between $30$ and $857\,\textrm{GHz}$. Built upon a Bayesian analysis framework, \emph{Planck's} foreground maps are created by fitting its observational data to parametric models \citep{planck_2015_X}. Since these models require an explicit modeling of foregrounds, they provide a useful contrast to our blind extractions.

\begin{figure}
\includegraphics[width=0.49\textwidth]{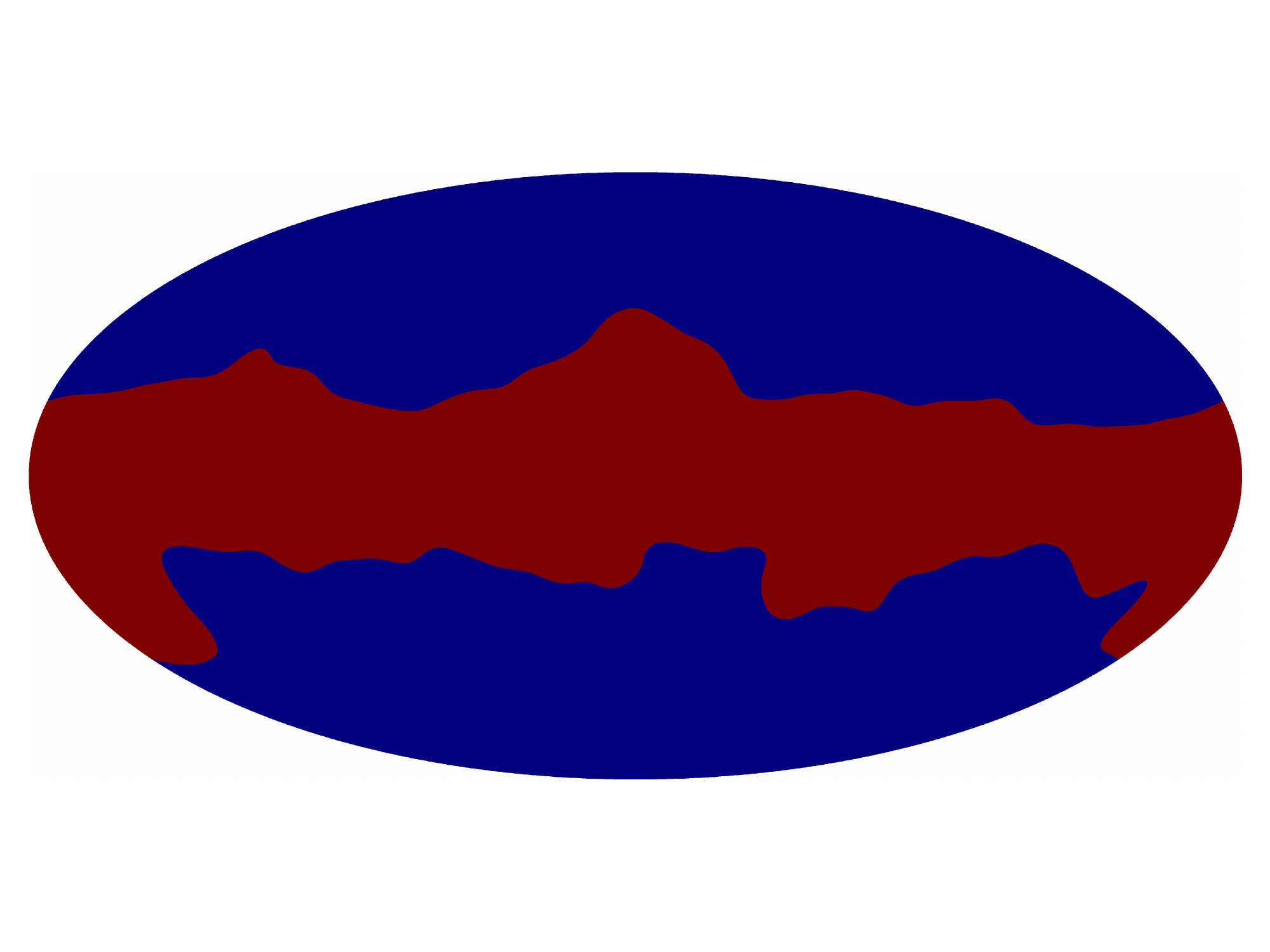}
\caption{``In-plane" (red region) and ``out-of-plane" (blue region) masks used for the comparison of our physical foreground components with those from \emph{Planck}.}
\label{fig:fgMasks}
\end{figure}

To compare the foreground components extracted from the GSM and their counterparts from \emph{Planck}, we calculate a normalized angular cross-power spectrum $R_{l}^\textrm{Planck, GSM}$
\begin{equation}
\label{eq:CrossCorr}
R_{\ell}^\textrm{Planck, GSM}= \frac{C_{\ell}^\textrm{Planck, GSM}}{\sqrt{C_{\ell}^\textrm{Planck} C_{\ell}^\textrm{GSM}}}
\end{equation}
where $C_{\ell}^{X,Y} = \frac{1}{2l +1}\sum_{m} a_{\ell m}^{X} a_{\ell m}^{Y*}$ with $a_{\ell m}$ denoting a spherical harmonic expansion coefficient for spherical harmonic indices $\ell$ and $m$. We choose to normalize our cross-powers to minimize effects due to differences in instrumental properties between \emph{Planck} and the wide variety of data used to construct the GSM. We also choose to split our comparison into regions in and out of the Galactic plane, defined by the map shown in Figure \ref{fig:fgMasks}. Note that \emph{Planck} uses a one-component gray-body thermal dust model, and for our comparison we correlate both dust components with the single dust map from \emph{Planck}.

\begin{figure*}
\includegraphics[width=1\textwidth,trim=2.0cm 3.2cm 2.8cm 3.5cm,clip]{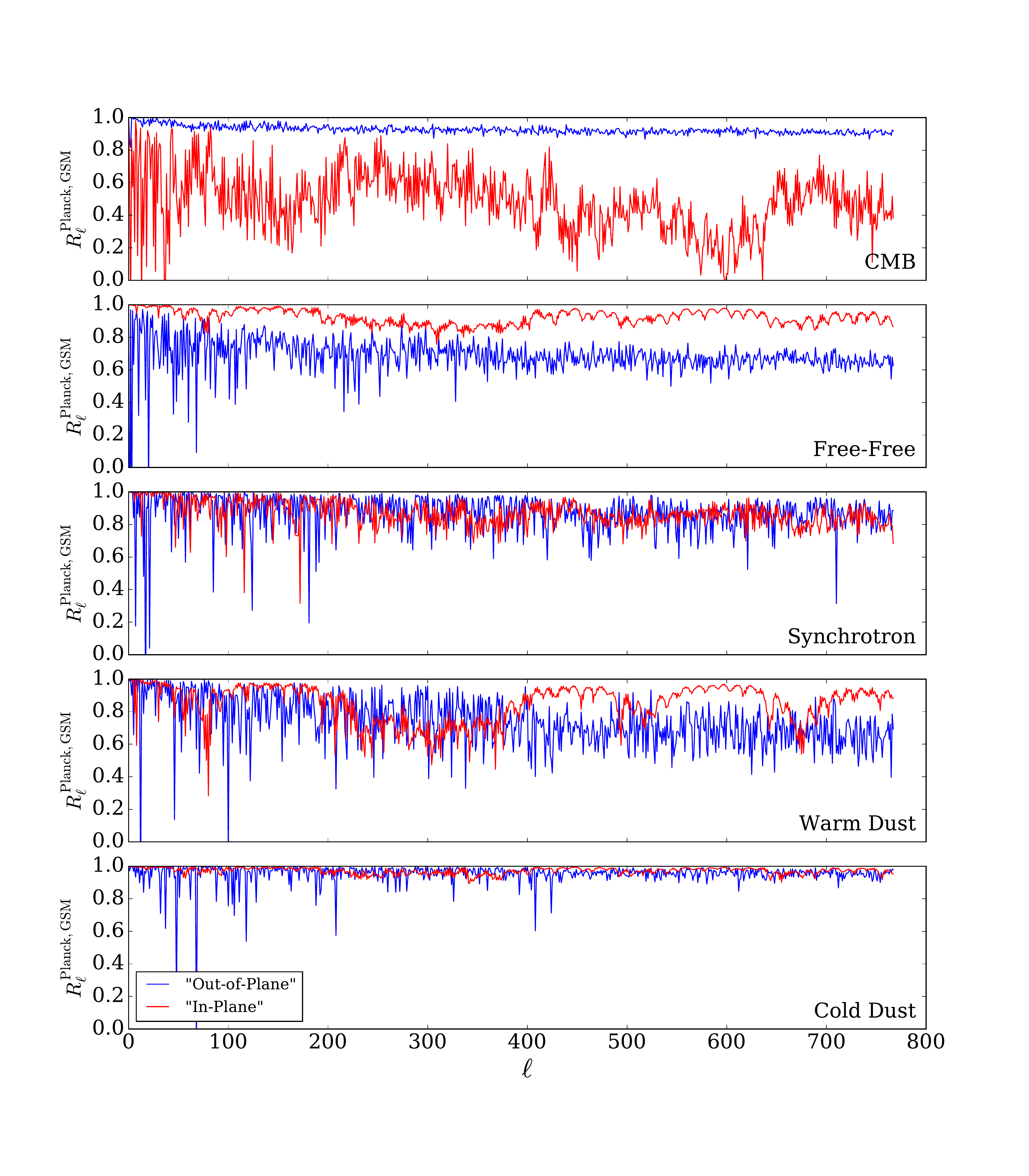}
\caption{The normalized cross-correlation $R_{\ell}^\textrm{Planck, GSM}$ from Eq. \eqref{eq:CrossCorr}, for the CMB, free-free, synchrotron, and dust components. Overall, there is a reasonable correlation between our extracted foreground components and those from \emph{Planck} throughout all angular scales on most of components. Red lines denote the cross-correlation inside the Galactic plane, while blue lines show the cross-correlation outside of the Galactic plane. Our CMB map is seen to be only weakly correlated with the corresponding \emph{Planck} map in the Galactic plane, while highly correlated outside the plane.}
\label{fig:CrossCorr}
\end{figure*}

Figure \ref{fig:CrossCorr} shows the resulting cross-power spectra. The extracted CMB maps are most strongly correlated outside of the Galactic plane, where possible confusion with foregrounds (particularly with our blind extraction method) is least likely. The lower correlation within the Galactic plane suggest that there is indeed some leakage from the foreground components. For most of our foreground components, we generally find strong cross-correlations between our maps and those from \emph{Planck}, with only a very weak dependence on angular scale. These trends are relatively robust to the precise mask that we use to define the regions inside and outside the plane (for example, one could have chosen a mask that is morphologically more similar to free-free emission than our current mask, which is more similar to the patterns of dust emission). There are, however, two minor exceptions to this: we find the out-of-plane curves for free-free and warm dust to shift up and down slightly depending on the exact mask used. This is perhaps not surprising, given that warm dust and free-free are two of the components that are more compactly confined to the Galactic plane, and thus the most sensitive to the definition of the ``plane". This does, however, suggest a Galactic latitude-dependence on the accuracy of our modeling procedures. A detailed examination of such dependence is beyond the scope of this paper, and investigations are currently underway to incorporate in our modeling the fact that the Galactic plane is a more physically complex region than the Galactic poles. Finally, we find that our sixth component, HI+other, is not highly correlated with any of \emph{Planck's} foreground products.

Ultimately, it is perhaps unsurprising that there is a reasonable correlation between our extracted components and those from \emph{Planck}. After all, \emph{Planck} maps were used to form the GSM in the first place. Thus, the exercise performed here is more a self-consistent demonstration of the ability of our blind method to identify independent foreground components. Our method may thus be helpful for future iterations of the GSM, where it may be used to detect the presence of new foreground components demanded by better data, prior to the availability of a physical model.

%%%%%%%%%%%%%%%%%%%%%%%%%%%%%
%%%%%%%%%%%%%%%%%%%%%%%%%%%%%
%%%%%%%%%%%%%%%%%%%%%%%%%%%%%
\section{Summary and Outlook}\label{secsummarygsm2}
We have presented an improved algorithm that builds upon the original PCA algorithm, allowing us to include many more incomplete sky maps across a larger frequency range. We have presented an improved GSM, with expected predictive accuracy between 5\% and 15\% for most frequencies, and around 2\% near CMB frequencies, with overall amplitude offsets less than 15\%. It is worth pointing out that these accuracy estimations are based on RMS values over the whole sky, so for a particular sky direction, there can be higher errors. We have also presented a blind component separation technique, which identifies five physical components that agree very well with existing physical models. Lastly, we also create a high resolution GSM, with $56'$ resolution at frequencies below 10\,GHz, and $24'$ resolution above 10\,GHz.

There are two ways to further improve the GSM in the future: adding more data sets and improving the algorithm. On the data side, many high quality maps will become available in the near future, such as \textit{CBASS} \citep{CBASS1, CBASS2}, \textit{SPASS} \citep{SPASS}, \textit{GMIMS} \citep{GMIMS}, \textit{GEM} \citep{GEM}, and \textit{QUIJOTE} \citep{QUIJOTE}. Some of these upcoming maps will fill in the gap between 2.3\,GHz and 23\,GHz. In addition, in a separate paper \citep{zhengMM}, we present a new imaging method that allows existing low-frequency interferometers such as the \textit{MWA} to produce high quality foreground maps that cover nearly the full sky. Last but not least, we aim to produce a polarised GSM using existing and upcoming polarised sky maps. 

On the algorithmic side, we would like to include uncertainty information for all sky maps in the form of error covariance matrices of the maps into \refeq{eqWiter} and \refeq{eqMiter}. This would allow us to properly weigh each map rather than resorting to renormalizing all maps and treating them all equally. In addition, this will allow us to perform rigorous model selection in order to decide the optimal number of principal components. Last but not least, we are interested in improving the blind physical component separation procedure to learn more about the physical mechanisms that correspond to the components found in the GSM.
 
With the help of upcoming high-quality sky maps, the improved GSM will not only serve as a powerful predictive model for diffuse Galactic emission, but will also have the potential to uncover new physical mechanisms that contribute to this emission.

{\bf Acknowledgments:}
This work was supported by NSF grants AST-1105835 and AST-1440343. AL acknowledges support from the University of California Office of the President Multicampus Research Programs and Initiatives through award MR-15-328388, and from NASA through Hubble Fellowship grant HST-HF2-51363.001-A awarded by the Space Telescope Science Institute, which is operated by the Association of Universities for Research in Astronomy, Inc., for NASA, under contract NAS5-26555. The NASA funding for the eGSM is NASA grant no. NNX16AF55G. We acknowledge the use of the Legacy Archive for Microwave Background Data Analysis (LAMBDA), part of the High Energy Astrophysics Science Archive Center (HEASARC). HEASARC/LAMBDA is a service of the Astrophysics Science Division at the NASA Goddard Space Flight Center. This research has made use of the NASA/ IPAC Infrared Science Archive, which is operated by the Jet Propulsion Laboratory, California Institute of Technology, under contract with the National Aeronautics and Space Administration. We wish to thank Jacqueline Hewitt, Gianni Bernardi, Douglas Finkbeiner, Aaron M. Meisner for helpful comments and suggestions, and Angelica de Oliveira-Costa for pioneering the GSM approach. We also wish to thank our anonymous referee for extremely helpful and detailed feedback.

\bibliographystyle{mnras}
\bibliography{MMGSM2}

\begin{thebibliography}{}
\makeatletter
\relax
\def\mn@urlcharsother{\let\do\@makeother \do\$\do\&\do\#\do\^\do\_\do\%\do\~}
\def\mn@doi{\begingroup\mn@urlcharsother \@ifnextchar [ {\mn@doi@}
  {\mn@doi@[]}}
\def\mn@doi@[#1]#2{\def\@tempa{#1}\ifx\@tempa\@empty \href
  {http://dx.doi.org/#2} {doi:#2}\else \href {http://dx.doi.org/#2} {#1}\fi
  \endgroup}
\def\mn@eprint#1#2{\mn@eprint@#1:#2::\@nil}
\def\mn@eprint@arXiv#1{\href {http://arxiv.org/abs/#1} {{\tt arXiv:#1}}}
\def\mn@eprint@dblp#1{\href {http://dblp.uni-trier.de/rec/bibtex/#1.xml}
  {dblp:#1}}
\def\mn@eprint@#1:#2:#3:#4\@nil{\def\@tempa {#1}\def\@tempb {#2}\def\@tempc
  {#3}\ifx \@tempc \@empty \let \@tempc \@tempb \let \@tempb \@tempa \fi \ifx
  \@tempb \@empty \def\@tempb {arXiv}\fi \@ifundefined
  {mn@eprint@\@tempb}{\@tempb:\@tempc}{\expandafter \expandafter \csname
  mn@eprint@\@tempb\endcsname \expandafter{\@tempc}}}

\bibitem[\protect\citeauthoryear{{Ali} et~al.,}{{Ali}
  et~al.}{2015}]{PAPERpspec2}
{Ali} Z.~S.,  et~al., 2015, \mn@doi [\apj] {10.1088/0004-637X/809/1/61}, \href
  {http://adsabs.harvard.edu/abs/2015ApJ...809...61A} {809, 61}

\bibitem[\protect\citeauthoryear{{Alvarez}, {Aparici}, {May}  \&
  {Olmos}}{{Alvarez} et~al.}{1997}]{Alvarez1997MRAO}
{Alvarez} H.,  {Aparici} J.,  {May} J.,   {Olmos} F.,  1997, \mn@doi [\aaps]
  {10.1051/aas:1997196}, \href
  {http://adsabs.harvard.edu/abs/1997A%26AS..124..315A} {124, 315}

\bibitem[\protect\citeauthoryear{{Banday}, {Dickinson}, {Davies}, {Davis}  \&
  {G{\'o}rski}}{{Banday} et~al.}{2003}]{AME3}
{Banday} A.~J.,  {Dickinson} C.,  {Davies} R.~D.,  {Davis} R.~J.,
  {G{\'o}rski} K.~M.,  2003, \mn@doi [\mnras]
  {10.1046/j.1365-8711.2003.07008.x}, \href
  {http://adsabs.harvard.edu/abs/2003MNRAS.345..897B} {345, 897}

\bibitem[\protect\citeauthoryear{{Bennett} et~al.,}{{Bennett}
  et~al.}{2013}]{wmap9}
{Bennett} C.~L.,  et~al., 2013, \mn@doi [\apjs] {10.1088/0067-0049/208/2/20},
  \href {http://adsabs.harvard.edu/abs/2013ApJS..208...20B} {208, 20}

\bibitem[\protect\citeauthoryear{{Berkhuijsen}}{{Berkhuijsen}}{1971}]{Berkhuijsen1971}
{Berkhuijsen} E.~M.,  1971, \aap, \href
  {http://adsabs.harvard.edu/abs/1971A%26A....14..359B} {14, 359}

\bibitem[\protect\citeauthoryear{{Bonaldi}, {Ricciardi}, {Leach}, {Stivoli},
  {Baccigalupi}  \& {de Zotti}}{{Bonaldi} et~al.}{2007}]{AME8}
{Bonaldi} A.,  {Ricciardi} S.,  {Leach} S.,  {Stivoli} F.,  {Baccigalupi} C.,
  {de Zotti} G.,  2007, \mn@doi [\mnras] {10.1111/j.1365-2966.2007.12477.x},
  \href {http://adsabs.harvard.edu/abs/2007MNRAS.382.1791B} {382, 1791}

\bibitem[\protect\citeauthoryear{{Burke-Spolaor} \&
  {Bannister}}{{Burke-Spolaor} \& {Bannister}}{2014}]{Burke-Spolaor}
{Burke-Spolaor} S.,  {Bannister} K.~W.,  2014, \mn@doi [\apj]
  {10.1088/0004-637X/792/1/19}, \href
  {http://adsabs.harvard.edu/abs/2014ApJ...792...19B} {792, 19}

\bibitem[\protect\citeauthoryear{{Calabretta}, {Staveley-Smith}  \&
  {Barnes}}{{Calabretta} et~al.}{2014}]{Calabretta2014CHIPASS}
{Calabretta} M.~R.,  {Staveley-Smith} L.,   {Barnes} D.~G.,  2014, \mn@doi
  [\pasa] {10.1017/pasa.2013.36}, \href
  {http://adsabs.harvard.edu/abs/2014PASA...31....7C} {31, 7}

\bibitem[\protect\citeauthoryear{{Carretti}, {Gaensler}, {Staveley-Smith},
  {Haverkorn}, {Kesteven}, {Cortiglioni}, {Bernardi}  \& {Poppi}}{{Carretti}
  et~al.}{2009}]{SPASS}
{Carretti} E.,  {Gaensler} B.,  {Staveley-Smith} L.,  {Haverkorn} M.,
  {Kesteven} M.,  {Cortiglioni} S.,  {Bernardi} G.,   {Poppi} S.,  2009,
  {S-band Polarization All Sky Survey (S-PASS)}, ATNF Proposal

\bibitem[\protect\citeauthoryear{{Caswell}}{{Caswell}}{1976}]{Caswell1976DRAO}
{Caswell} J.~L.,  1976, \mnras, \href
  {http://cdsads.u-strasbg.fr/abs/1976MNRAS.177..601C} {177, 601}

\bibitem[\protect\citeauthoryear{{Datta}, {Bowman}  \& {Carilli}}{{Datta}
  et~al.}{2010}]{foreground1}
{Datta} A.,  {Bowman} J.~D.,   {Carilli} C.~L.,  2010, \mn@doi [\apj]
  {10.1088/0004-637X/724/1/526}, \href
  {http://adsabs.harvard.edu/abs/2010ApJ...724..526D} {724, 526}

\bibitem[\protect\citeauthoryear{{Davies}, {Watson}  \& {Gutierrez}}{{Davies}
  et~al.}{1996}]{Davies1996}
{Davies} R.~D.,  {Watson} R.~A.,   {Gutierrez} C.~M.,  1996, \mn@doi [\mnras]
  {10.1093/mnras/278.4.925}, \href
  {http://adsabs.harvard.edu/abs/1996MNRAS.278..925D} {278, 925}

\bibitem[\protect\citeauthoryear{{Davies}, {Dickinson}, {Banday}, {Jaffe},
  {G{\'o}rski}  \& {Davis}}{{Davies} et~al.}{2006}]{AME7}
{Davies} R.~D.,  {Dickinson} C.,  {Banday} A.~J.,  {Jaffe} T.~R.,  {G{\'o}rski}
  K.~M.,   {Davis} R.~J.,  2006, \mn@doi [\mnras]
  {10.1111/j.1365-2966.2006.10572.x}, \href
  {http://adsabs.harvard.edu/abs/2006MNRAS.370.1125D} {370, 1125}

\bibitem[\protect\citeauthoryear{{\lowercase{D}e Oliveira-Costa} \&
  {Tegmark}}{{\lowercase{D}e Oliveira-Costa} \&
  {Tegmark}}{1999}]{all_sorts_of_cmb_papers}
{\lowercase{D}e Oliveira-Costa} A.,  {Tegmark} M.,  eds, 1999, {Microwave
  Foregrounds}  Astronomical Society of the Pacific Conference Series Vol. 181

\bibitem[\protect\citeauthoryear{{\lowercase{D}e Oliveira-Costa}, {Tegmark}, {Davies},
  {Guti{\'e}rrez}, {Lasenby}, {Rebolo}  \& {Watson}}{{de Oliveira-Costa}
  et~al.}{2004}]{AME5}
{de Oliveira-Costa} A.,  {Tegmark} M.,  {Davies} R.~D.,  {Guti{\'e}rrez} C.~M.,
   {Lasenby} A.~N.,  {Rebolo} R.,   {Watson} R.~A.,  2004, \mn@doi [\apjl]
  {10.1086/421293}, \href {http://adsabs.harvard.edu/abs/2004ApJ...606L..89D}
  {606, L89}

\bibitem[\protect\citeauthoryear{{\lowercase{D}e Oliveira-Costa}, {Tegmark},
  {Gaensler}, {Jonas}, {Landecker}  \& {Reich}}{{\lowercase{D}e Oliveira-Costa}
  et~al.}{2008}]{GSM}
{\lowercase{D}e Oliveira-Costa} A.,  {Tegmark} M.,  {Gaensler} B.~M.,  {Jonas}
  J.,  {Landecker} T.~L.,   {Reich} P.,  2008, \mn@doi [MNRAS]
  {10.1111/j.1365-2966.2008.13376.x}, \href
  {http://adsabs.harvard.edu/abs/2008MNRAS.388..247D} {388, 247}

\bibitem[\protect\citeauthoryear{{Delabrouille} et~al.,}{{Delabrouille}
  et~al.}{2013}]{PSM}
{Delabrouille} J.,  et~al., 2013, \mn@doi [\aap] {10.1051/0004-6361/201220019},
  \href {http://adsabs.harvard.edu/abs/2013A%26A...553A..96D} {553, A96}

\bibitem[\protect\citeauthoryear{{Dickinson}, {Davies}  \& {Davis}}{{Dickinson}
  et~al.}{2003}]{freefreeDickinson}
{Dickinson} C.,  {Davies} R.~D.,   {Davis} R.~J.,  2003, \mn@doi [\mnras]
  {10.1046/j.1365-8711.2003.06439.x}, \href
  {http://adsabs.harvard.edu/abs/2003MNRAS.341..369D} {341, 369}

\bibitem[\protect\citeauthoryear{{Dillon} et~al.,}{{Dillon}
  et~al.}{2014}]{MWAJosh}
{Dillon} J.~S.,  et~al., 2014, \mn@doi [\prd] {10.1103/PhysRevD.89.023002},
  \href {http://adsabs.harvard.edu/abs/2014PhRvD..89b3002D} {89, 023002}

\bibitem[\protect\citeauthoryear{{Dillon} et~al.,}{{Dillon}
  et~al.}{2015}]{dillon_et_al2015}
{Dillon} J.~S.,  et~al., 2015, \mn@doi [\prd] {10.1103/PhysRevD.91.123011},
  \href {http://adsabs.harvard.edu/abs/2015PhRvD..91l3011D} {91, 123011}

\bibitem[\protect\citeauthoryear{{Dobler} \& {Finkbeiner}}{{Dobler} \&
  {Finkbeiner}}{2008}]{AME9}
{Dobler} G.,  {Finkbeiner} D.~P.,  2008, \mn@doi [\apj] {10.1086/587862}, \href
  {http://adsabs.harvard.edu/abs/2008ApJ...680.1222D} {680, 1222}

\bibitem[\protect\citeauthoryear{{Doi} et~al.,}{{Doi}
  et~al.}{2015}]{Doi2015AKARI}
{Doi} Y.,  et~al., 2015, \mn@doi [\pasj] {10.1093/pasj/psv022}, \href
  {http://adsabs.harvard.edu/abs/2015PASJ...67...50D} {67, 50}

\bibitem[\protect\citeauthoryear{{Finkbeiner}}{{Finkbeiner}}{2003}]{Halpha}
{Finkbeiner} D.~P.,  2003, \mn@doi [\apjs] {10.1086/374411}, \href
  {http://adsabs.harvard.edu/abs/2003ApJS..146..407F} {146, 407}

\bibitem[\protect\citeauthoryear{{Finkbeiner}}{{Finkbeiner}}{2004}]{AME6}
{Finkbeiner} D.~P.,  2004, \mn@doi [\apj] {10.1086/423482}, \href
  {http://adsabs.harvard.edu/abs/2004ApJ...614..186F} {614, 186}

\bibitem[\protect\citeauthoryear{{Finkbeiner}, {Davis}  \&
  {Schlegel}}{{Finkbeiner} et~al.}{1999}]{dust1}
{Finkbeiner} D.~P.,  {Davis} M.,   {Schlegel} D.~J.,  1999, \mn@doi [\apj]
  {10.1086/307852}, \href {http://adsabs.harvard.edu/abs/1999ApJ...524..867F}
  {524, 867}

\bibitem[\protect\citeauthoryear{{Fixsen}}{{Fixsen}}{2009}]{cmbtemp}
{Fixsen} D.~J.,  2009, \mn@doi [\apj] {10.1088/0004-637X/707/2/916}, \href
  {http://adsabs.harvard.edu/abs/2009ApJ...707..916F} {707, 916}

\bibitem[\protect\citeauthoryear{{Furlanetto}, {Oh}  \& {Briggs}}{{Furlanetto}
  et~al.}{2006}]{FurlanettoReview}
{Furlanetto} S.~R.,  {Oh} S.~P.,   {Briggs} F.~H.,  2006, \mn@doi [\physrep]
  {10.1016/j.physrep.2006.08.002}, \href
  {http://adsabs.harvard.edu/abs/2006PhR...433..181F} {433, 181}

\bibitem[\protect\citeauthoryear{{G{\'e}nova-Santos}
  et~al.,}{{G{\'e}nova-Santos} et~al.}{2015}]{QUIJOTE}
{G{\'e}nova-Santos} R.,  et~al., 2015, \mn@doi [\mnras]
  {10.1093/mnras/stv1405}, \href
  {http://adsabs.harvard.edu/abs/2015MNRAS.452.4169G} {452, 4169}

\bibitem[\protect\citeauthoryear{{Gold} et~al.,}{{Gold} et~al.}{2011}]{AME12}
{Gold} B.,  et~al., 2011, \mn@doi [\apjs] {10.1088/0067-0049/192/2/15}, \href
  {http://adsabs.harvard.edu/abs/2011ApJS..192...15G} {192, 15}

\bibitem[\protect\citeauthoryear{{G{\'o}rski}, {Hivon}, {Banday}, {Wandelt},
  {Hansen}, {Reinecke}  \& {Bartelmann}}{{G{\'o}rski} et~al.}{2005}]{HEALPIX}
{G{\'o}rski} K.~M.,  {Hivon} E.,  {Banday} A.~J.,  {Wandelt} B.~D.,  {Hansen}
  F.~K.,  {Reinecke} M.,   {Bartelmann} M.,  2005, \mn@doi [\apj]
  {10.1086/427976}, \href {http://adsabs.harvard.edu/abs/2005ApJ...622..759G}
  {622, 759}

\bibitem[\protect\citeauthoryear{{Haslam}, {Klein}, {Salter}, {Stoffel},
  {Wilson}, {Cleary}, {Cooke}  \& {Thomasson}}{{Haslam}
  et~al.}{1981}]{Haslam1981}
{Haslam} C.~G.~T.,  {Klein} U.,  {Salter} C.~J.,  {Stoffel} H.,  {Wilson}
  W.~E.,  {Cleary} M.~N.,  {Cooke} D.~J.,   {Thomasson} P.,  1981, \aap, \href
  {http://adsabs.harvard.edu/abs/1981A%26A...100..209H} {100, 209}

\bibitem[\protect\citeauthoryear{{Haslam}, {Salter}, {Stoffel}  \&
  {Wilson}}{{Haslam} et~al.}{1982}]{Haslam1982}
{Haslam} C.~G.~T.,  {Salter} C.~J.,  {Stoffel} H.,   {Wilson} W.~E.,  1982,
  \aaps, \href {http://adsabs.harvard.edu/abs/1982A%26AS...47....1H} {47, 1}

\bibitem[\protect\citeauthoryear{{Hinshaw} et~al.,}{{Hinshaw}
  et~al.}{2009}]{WMAP5year}
{Hinshaw} G.,  et~al., 2009, \mn@doi [\apjs] {10.1088/0067-0049/180/2/225},
  \href {http://adsabs.harvard.edu/abs/2009ApJS..180..225H} {180, 225}

\bibitem[\protect\citeauthoryear{{Irfan} et~al.,}{{Irfan}
  et~al.}{2015}]{CBASS2}
{Irfan} M.~O.,  et~al., 2015, \mn@doi [\mnras] {10.1093/mnras/stv212}, \href
  {http://adsabs.harvard.edu/abs/2015MNRAS.448.3572I} {448, 3572}

\bibitem[\protect\citeauthoryear{{Jonas}, {Baart}  \& {Nicolson}}{{Jonas}
  et~al.}{1998}]{Jonas1998Rhodes}
{Jonas} J.~L.,  {Baart} E.~E.,   {Nicolson} G.~D.,  1998, \mn@doi [\mnras]
  {10.1046/j.1365-8711.1998.01367.x}, \href
  {http://adsabs.harvard.edu/abs/1998MNRAS.297..977J} {297, 977}

\bibitem[\protect\citeauthoryear{King et~al.,}{King et~al.}{2010}]{CBASS1}
King O.~G.,  et~al., 2010. pp 77411I--77411I--10, \mn@doi{10.1117/12.858011},
  \url {http://dx.doi.org/10.1117/12.858011}

\bibitem[\protect\citeauthoryear{{Kogut}}{{Kogut}}{1996}]{AME1}
{Kogut} A.,  1996, in American Astronomical Society Meeting Abstracts. p.~1295

\bibitem[\protect\citeauthoryear{{Lagache}}{{Lagache}}{2003}]{AME4}
{Lagache} G.,  2003, \mn@doi [\aap] {10.1051/0004-6361:20030545}, \href
  {http://adsabs.harvard.edu/abs/2003A%26A...405..813L} {405, 813}

\bibitem[\protect\citeauthoryear{{Landecker} \& {Wielebinski}}{{Landecker} \&
  {Wielebinski}}{1970}]{Landecker1970Parkes}
{Landecker} T.~L.,  {Wielebinski} R.,  1970, Australian Journal of Physics
  Astrophysical Supplement, \href
  {http://cdsads.u-strasbg.fr/abs/1970AuJPA..16....1L} {16, 1}

\bibitem[\protect\citeauthoryear{{Lawson}, {Mayer}, {Osborne}  \&
  {Parkinson}}{{Lawson} et~al.}{1987}]{Lawson1987}
{Lawson} K.~D.,  {Mayer} C.~J.,  {Osborne} J.~L.,   {Parkinson} M.~L.,  1987,
  \mn@doi [\mnras] {10.1093/mnras/225.2.307}, \href
  {http://adsabs.harvard.edu/abs/1987MNRAS.225..307L} {225, 307}

\bibitem[\protect\citeauthoryear{{Leitch}, {Readhead}, {Pearson}  \&
  {Myers}}{{Leitch} et~al.}{1997}]{AME2}
{Leitch} E.~M.,  {Readhead} A.~C.~S.,  {Pearson} T.~J.,   {Myers} S.~T.,  1997,
  \mn@doi [\apjl] {10.1086/310823}, \href
  {http://adsabs.harvard.edu/abs/1997ApJ...486L..23L} {486, L23}

\bibitem[\protect\citeauthoryear{{Liu}, {Parsons}  \& {Trott}}{{Liu}
  et~al.}{2014a}]{liu_et_al2014a}
{Liu} A.,  {Parsons} A.~R.,   {Trott} C.~M.,  2014a, \mn@doi [\prd]
  {10.1103/PhysRevD.90.023018}, \href
  {http://adsabs.harvard.edu/abs/2014PhRvD..90b3018L} {90, 023018}

\bibitem[\protect\citeauthoryear{{Liu}, {Parsons}  \& {Trott}}{{Liu}
  et~al.}{2014b}]{liu_et_al2014b}
{Liu} A.,  {Parsons} A.~R.,   {Trott} C.~M.,  2014b, \mn@doi [\prd]
  {10.1103/PhysRevD.90.023019}, \href
  {http://adsabs.harvard.edu/abs/2014PhRvD..90b3019L} {90, 023019}

\bibitem[\protect\citeauthoryear{{Maeda}, {Alvarez}, {Aparici}, {May}  \&
  {Reich}}{{Maeda} et~al.}{1999}]{Maeda1999JMUAR}
{Maeda} K.,  {Alvarez} H.,  {Aparici} J.,  {May} J.,   {Reich} P.,  1999,
  \mn@doi [\aaps] {10.1051/aas:1999413}, \href
  {http://adsabs.harvard.edu/abs/1999A%26AS..140..145M} {140, 145}

\bibitem[\protect\citeauthoryear{{Meisner} \& {Finkbeiner}}{{Meisner} \&
  {Finkbeiner}}{2015}]{dust2}
{Meisner} A.~M.,  {Finkbeiner} D.~P.,  2015, \mn@doi [\apj]
  {10.1088/0004-637X/798/2/88}, \href
  {http://adsabs.harvard.edu/abs/2015ApJ...798...88M} {798, 88}

\bibitem[\protect\citeauthoryear{{Miville-Desch{\^e}nes} \&
  {Lagache}}{{Miville-Desch{\^e}nes} \& {Lagache}}{2005}]{Miville2005IRIS}
{Miville-Desch{\^e}nes} M.-A.,  {Lagache} G.,  2005, \mn@doi [\apjs]
  {10.1086/427938}, \href {http://adsabs.harvard.edu/abs/2005ApJS..157..302M}
  {157, 302}

\bibitem[\protect\citeauthoryear{{Miville-Desch{\^e}nes}, {Ysard}, {Lavabre},
  {Ponthieu}, {Mac{\'{\i}}as-P{\'e}rez}, {Aumont}  \&
  {Bernard}}{{Miville-Desch{\^e}nes} et~al.}{2008}]{AME10}
{Miville-Desch{\^e}nes} M.-A.,  {Ysard} N.,  {Lavabre} A.,  {Ponthieu} N.,
  {Mac{\'{\i}}as-P{\'e}rez} J.~F.,  {Aumont} J.,   {Bernard} J.~P.,  2008,
  \mn@doi [\aap] {10.1051/0004-6361:200809484}, \href
  {http://adsabs.harvard.edu/abs/2008A%26A...490.1093M} {490, 1093}

\bibitem[\protect\citeauthoryear{{Morales} \& {Wyithe}}{{Morales} \&
  {Wyithe}}{2010}]{miguelreview}
{Morales} M.~F.,  {Wyithe} J.~S.~B.,  2010, \mn@doi [\araa]
  {10.1146/annurev-astro-081309-130936}, \href
  {http://adsabs.harvard.edu/abs/2010ARA%26A..48..127M} {48, 127}

\bibitem[\protect\citeauthoryear{{Morales}, {Hazelton}, {Sullivan}  \&
  {Beardsley}}{{Morales} et~al.}{2012}]{foreground4}
{Morales} M.~F.,  {Hazelton} B.,  {Sullivan} I.,   {Beardsley} A.,  2012,
  \mn@doi [\apj] {10.1088/0004-637X/752/2/137}, \href
  {http://adsabs.harvard.edu/abs/2012ApJ...752..137M} {752, 137}

\bibitem[\protect\citeauthoryear{{Mozdzen}, {Bowman}, {Monsalve}  \&
  {Rogers}}{{Mozdzen} et~al.}{2016}]{foreground9}
{Mozdzen} T.~J.,  {Bowman} J.~D.,  {Monsalve} R.~A.,   {Rogers} A.~E.~E.,
  2016, \mn@doi [\mnras] {10.1093/mnras/stv2601}, \href
  {http://adsabs.harvard.edu/abs/2016MNRAS.455.3890M} {455, 3890}

\bibitem[\protect\citeauthoryear{{Novaes}, {Bernui}, {Ferreira}  \&
  {Wuensche}}{{Novaes} et~al.}{2015}]{Novaes2015}
{Novaes} C.~P.,  {Bernui} A.,  {Ferreira} I.~S.,   {Wuensche} C.~A.,  2015,
  \mn@doi [\jcap] {10.1088/1475-7516/2015/09/064}, \href
  {http://adsabs.harvard.edu/abs/2015JCAP...09..064N} {9, 064}

\bibitem[\protect\citeauthoryear{{Parsons} et~al.,}{{Parsons}
  et~al.}{2010}]{PAPER}
{Parsons} A.~R.,  et~al., 2010, \mn@doi [\aj] {10.1088/0004-6256/139/4/1468},
  \href {http://adsabs.harvard.edu/abs/2010AJ....139.1468P} {139, 1468}

\bibitem[\protect\citeauthoryear{{Parsons}, {Pober}, {Aguirre}, {Carilli},
  {Jacobs}  \& {Moore}}{{Parsons} et~al.}{2012}]{foreground3}
{Parsons} A.~R.,  {Pober} J.~C.,  {Aguirre} J.~E.,  {Carilli} C.~L.,  {Jacobs}
  D.~C.,   {Moore} D.~F.,  2012, \mn@doi [\apj] {10.1088/0004-637X/756/2/165},
  \href {http://adsabs.harvard.edu/abs/2012ApJ...756..165P} {756, 165}

\bibitem[\protect\citeauthoryear{{Parsons} et~al.,}{{Parsons}
  et~al.}{2014}]{PAPERpspec}
{Parsons} A.~R.,  et~al., 2014, \mn@doi [\apj] {10.1088/0004-637X/788/2/106},
  \href {http://adsabs.harvard.edu/abs/2014ApJ...788..106P} {788, 106}

\bibitem[\protect\citeauthoryear{{Planck Collaboration} et~al.,}{{Planck
  Collaboration} et~al.}{2015a}]{planck2015}
{Planck Collaboration} et~al., 2015a, preprint, \href
  {http://adsabs.harvard.edu/abs/2015arXiv150201582P} {} (\mn@eprint {arXiv}
  {1502.01582})

\bibitem[\protect\citeauthoryear{{Planck Collaboration} et~al.,}{{Planck
  Collaboration} et~al.}{2015b}]{planck_2015_X}
{Planck Collaboration} et~al., 2015b, preprint, \href
  {http://adsabs.harvard.edu/abs/2015arXiv150201588P} {} (\mn@eprint {arXiv}
  {1502.01588})

\bibitem[\protect\citeauthoryear{{Platania}, {Burigana}, {Maino}, {Caserini},
  {Bersanelli}, {Cappellini}  \& {Mennella}}{{Platania}
  et~al.}{2003}]{platania2003}
{Platania} P.,  {Burigana} C.,  {Maino} D.,  {Caserini} E.,  {Bersanelli} M.,
  {Cappellini} B.,   {Mennella} A.,  2003, \mn@doi [\aap]
  {10.1051/0004-6361:20031125}, \href
  {http://adsabs.harvard.edu/abs/2003A%26A...410..847P} {410, 847}

\bibitem[\protect\citeauthoryear{{Pober}}{{Pober}}{2015}]{foreground7}
{Pober} J.~C.,  2015, \mn@doi [\mnras] {10.1093/mnras/stu2575}, \href
  {http://adsabs.harvard.edu/abs/2015MNRAS.447.1705P} {447, 1705}

\bibitem[\protect\citeauthoryear{{Pober} et~al.,}{{Pober}
  et~al.}{2013}]{pober_etal2013b}
{Pober} J.~C.,  et~al., 2013, \mn@doi [\apjl] {10.1088/2041-8205/768/2/L36},
  \href {http://adsabs.harvard.edu/abs/2013ApJ...768L..36P} {768, L36}

\bibitem[\protect\citeauthoryear{{Pober} et~al.,}{{Pober}
  et~al.}{2016}]{foreground10}
{Pober} J.~C.,  et~al., 2016, \mn@doi [\apj] {10.3847/0004-637X/819/1/8}, \href
  {http://adsabs.harvard.edu/abs/2016ApJ...819....8P} {819, 8}

\bibitem[\protect\citeauthoryear{{Reich}}{{Reich}}{1982}]{Reich1982Stockert}
{Reich} W.,  1982, \aaps, \href
  {http://cdsads.u-strasbg.fr/abs/1982A%26AS...48..219R} {48, 219}

\bibitem[\protect\citeauthoryear{{Reich} \& {Reich}}{{Reich} \&
  {Reich}}{1986}]{Reich1986Stockert}
{Reich} P.,  {Reich} W.,  1986, \aaps, \href
  {http://cdsads.u-strasbg.fr/abs/1986A%26AS...63..205R} {63, 205}

\bibitem[\protect\citeauthoryear{{Reich} \& {Reich}}{{Reich} \&
  {Reich}}{1988}]{reichreich1988}
{Reich} P.,  {Reich} W.,  1988, \aaps, \href
  {http://adsabs.harvard.edu/abs/1988A%26AS...74....7R} {74, 7}

\bibitem[\protect\citeauthoryear{{Reich}, {Testori}  \& {Reich}}{{Reich}
  et~al.}{2001}]{Reich2001Elisa}
{Reich} P.,  {Testori} J.~C.,   {Reich} W.,  2001, \mn@doi [\aap]
  {10.1051/0004-6361:20011000}, \href
  {http://cdsads.u-strasbg.fr/abs/2001A%26A...376..861R} {376, 861}

\bibitem[\protect\citeauthoryear{{Remazeilles}, {Dickinson}, {Banday},
  {Bigot-Sazy}  \& {Ghosh}}{{Remazeilles} et~al.}{2015}]{Remazeilles2015Haslam}
{Remazeilles} M.,  {Dickinson} C.,  {Banday} A.~J.,  {Bigot-Sazy} M.-A.,
  {Ghosh} T.,  2015, \mn@doi [\mnras] {10.1093/mnras/stv1274}, \href
  {http://adsabs.harvard.edu/abs/2015MNRAS.451.4311R} {451, 4311}

\bibitem[\protect\citeauthoryear{{Roger}, {Costain}, {Landecker}  \&
  {Swerdlyk}}{{Roger} et~al.}{1999}]{Roger1999DRAO}
{Roger} R.~S.,  {Costain} C.~H.,  {Landecker} T.~L.,   {Swerdlyk} C.~M.,  1999,
  \mn@doi [\aaps] {10.1051/aas:1999239}, \href
  {http://cdsads.u-strasbg.fr/abs/1999A%26AS..137....7R} {137, 7}

\bibitem[\protect\citeauthoryear{{Rogers} \& {Bowman}}{{Rogers} \&
  {Bowman}}{2008}]{EDGESspectralindex}
{Rogers} A.~E.~E.,  {Bowman} J.~D.,  2008, \mn@doi [\aj]
  {10.1088/0004-6256/136/2/641}, \href
  {http://adsabs.harvard.edu/abs/2008AJ....136..641R} {136, 641}

\bibitem[\protect\citeauthoryear{{Spekkens}, {Mason}, {Aguirre}  \&
  {Nhan}}{{Spekkens} et~al.}{2013}]{Spekkens2013}
{Spekkens} K.,  {Mason} B.~S.,  {Aguirre} J.~E.,   {Nhan} B.,  2013, \mn@doi
  [\apj] {10.1088/0004-637X/773/1/61}, \href
  {http://adsabs.harvard.edu/abs/2013ApJ...773...61S} {773, 61}

\bibitem[\protect\citeauthoryear{{Tello} et~al.,}{{Tello} et~al.}{2013}]{GEM}
{Tello} C.,  et~al., 2013, \mn@doi [\aap] {10.1051/0004-6361/20079306}, \href
  {http://adsabs.harvard.edu/abs/2013A%26A...556A...1T} {556, A1}

\bibitem[\protect\citeauthoryear{{Thyagarajan} et~al.,}{{Thyagarajan}
  et~al.}{2013}]{foreground5}
{Thyagarajan} N.,  et~al., 2013, \mn@doi [\apj] {10.1088/0004-637X/776/1/6},
  \href {http://adsabs.harvard.edu/abs/2013ApJ...776....6T} {776, 6}

\bibitem[\protect\citeauthoryear{{Thyagarajan} et~al.,}{{Thyagarajan}
  et~al.}{2015}]{foreground8}
{Thyagarajan} N.,  et~al., 2015, \mn@doi [\apj] {10.1088/0004-637X/804/1/14},
  \href {http://adsabs.harvard.edu/abs/2015ApJ...804...14T} {804, 14}

\bibitem[\protect\citeauthoryear{{Tingay} et~al.,}{{Tingay} et~al.}{2013}]{MWA}
{Tingay} S.~J.,  et~al., 2013, \mn@doi [\pasa] {10.1017/pasa.2012.007}, \href
  {http://adsabs.harvard.edu/abs/2013PASA...30....7T} {30, 7}

\bibitem[\protect\citeauthoryear{{Trott}, {Wayth}  \& {Tingay}}{{Trott}
  et~al.}{2012}]{foreground6}
{Trott} C.~M.,  {Wayth} R.~B.,   {Tingay} S.~J.,  2012, \mn@doi [\apj]
  {10.1088/0004-637X/757/1/101}, \href
  {http://adsabs.harvard.edu/abs/2012ApJ...757..101T} {757, 101}

\bibitem[\protect\citeauthoryear{{Vedantham}, {Udaya Shankar}  \&
  {Subrahmanyan}}{{Vedantham} et~al.}{2012}]{foreground2}
{Vedantham} H.,  {Udaya Shankar} N.,   {Subrahmanyan} R.,  2012, \mn@doi [\apj]
  {10.1088/0004-637X/745/2/176}, \href
  {http://adsabs.harvard.edu/abs/2012ApJ...745..176V} {745, 176}

\bibitem[\protect\citeauthoryear{{Wolleben} et~al.,}{{Wolleben}
  et~al.}{2009}]{GMIMS}
{Wolleben} M.,  et~al., 2009, in {Strassmeier} K.~G.,  {Kosovichev} A.~G.,
  {Beckman} J.~E.,  eds,  IAU Symposium Vol. 259, Cosmic Magnetic Fields: From
  Planets, to Stars and Galaxies. pp 89--90 (\mn@eprint {arXiv} {0812.2450}),
  \mn@doi{10.1017/S1743921309030117}

\bibitem[\protect\citeauthoryear{{Ysard}, {Miville-Desch{\^e}nes}  \&
  {Verstraete}}{{Ysard} et~al.}{2010}]{AME11}
{Ysard} N.,  {Miville-Desch{\^e}nes} M.~A.,   {Verstraete} L.,  2010, \mn@doi
  [\aap] {10.1051/0004-6361/200912715}, \href
  {http://adsabs.harvard.edu/abs/2010A%26A...509L...1Y} {509, L1}

\bibitem[\protect\citeauthoryear{{Zheng} et~al.,}{{Zheng}
  et~al.}{2016}]{zhengMM}
{Zheng} H.,  et~al., 2016, preprint, \href
  {http://adsabs.harvard.edu/abs/2016arXiv160503980Z} {} (\mn@eprint {arXiv}
  {1605.03980})



\makeatother
\end{thebibliography}
\end{document}